\documentclass[singlecolumn]{revtex4}
\usepackage{color,epsfig}
\usepackage{bm}
\usepackage{amsmath,amsfonts,amssymb,bm}
\usepackage{hyperref}
\usepackage{subfigure}

\newcommand{\bea}{\begin{eqnarray}}
\newcommand{\eea}{\end{eqnarray}}
\newcommand{\be}{\begin{equation}}
\newcommand{\ee}{\end{equation}}

\newcommand{\sgn}{{\rm sign}}
\def\etal{{\it et al.}}
\renewcommand\vec{\bm}

\begin{document}
\title{Analytic approach to magneto-strain tuning of electronic transport through a graphene nanobubble: Perspectives for a strain sensor}

\author{Enrique Mu\~{n}oz}
\email{munozt@fis.puc.cl}
\affiliation{Facultad de F\'isica, Pontificia Universidad Cat\'olica de Chile, Vicu\~{n}a Mackenna 4860, Santiago, Chile}
\author{Rodrigo Soto-Garrido}
\affiliation{Facultad de Ingenier\'ia y Tecnolog\'ia, Universidad San Sebasti\'an, Bellavista 7, Santiago 8420524, Chile} 

\date{\today}

\begin{abstract} We consider the scattering of Dirac particles in graphene due to the superposition of an external magnetic field and mechanical strain. As a model for a graphene nanobubble, we find exact analytical solutions for single-particle states inside and outside a circular region submitted to the fields. Finally, we obtain analytical expressions for the scattering cross-section, as well
as for the Landauer current through the circular region.  Our results provide a fully-analytical treatment for electronic transport through a graphene nanobubble, showing that a combination of a physical magnetic field and strain leads to valley polarization and filtering of the electronic current. Moreover, our analytical model provides an explicit metrology principle to measure strain by performing conductance experiments
under a controlled magnetic field imposed over the sample.
\end{abstract}

\maketitle

\section{Introduction}
\label{sec:Intro}
Graphene is an allotrope of carbon in the form of an atomic monolayer, arranged as a honeycomb lattice
with $C_{6v} = Z_2\otimes C_{3v}$ symmetry.
It is then mathematically described as a superposition of two Bravais lattices with $C_{3v}$ symmetry, usually denoted as A and B sub lattices, respectively \cite{Wallace_47,Geim_11,CastroNeto_09,Munoz_16}. As a consequence tight-binding, as well as ab-initio band structure 
calculations, show that the energy spectrum possesses linear dispersion in the vicinity of two non-equivalent, so called Dirac points (or valleys) in reciprocal space \cite{Geim_11,CastroNeto_09,Novoselov_05,Peres_10}. 
This particular feature allows for the description of graphene electronic properties
in terms of an effective Dirac Hamiltonian, whose eigenstates are given by two-component spinors,
where a pseudo-spin property emerges as a consequence of the two sub-lattices\cite{Peres_10,Novoselov_05,CastroNeto_09,Kim_08,Munoz_16}.
Those states exhibit pseudo-relativistic properties, such as relativistic Landau levels
in the presence of an external magnetic field \cite{Goerbig_11,CastroNeto_09,DasSarma_11}, where
the two Dirac points are connected by time-reversal symmetry, and hence the two valleys are degenerate \cite{Goerbig_11}. Perhaps an even more interesting feature arises under the presence
of mechanical strain. Within the Dirac approximation, strain enters as a gauge field whose
curl represents a pseudo-magnetic field that reverses sign at each Dirac point, thus breaking
the valley symmetry \cite{Vozmediano_10, Guinea_10, Guinea_Geim_10, Low_010, deJuan_13,Guinea_08,Munoz_15,Amorim_016}. Conductance experiments have shown the emergence of pseudo-relativistic Landau levels in the presence of strain solely, thus suggesting that the magnitude of
the associated pseudo-magnetic fields can reach over 100 Tesla for a small nano-bubble \cite{Levy_10,Klimov_012} or ridge \cite{Yan_012} in graphene.
From the theoretical perspective, strain-induced gauge fields have been incorporated into
extended Dirac Hamiltonians that involve the simultaneous description of both non-equivalent
Dirac cones \cite{Vozmediano_10,deJuan_13,deJuan_12,Gopalakrishnan2012,Munoz_15,Amorim_016}. Other physical
effects, such as charge density waves, can also be included in the form of generalized SU(2) gauge fields\cite{Gopalakrishnan2012}. 

Arbitrary strain patterns
generate inhomogeneous pseudo-magnetic fields, and in an experimental sample it is difficult
to characterize with nanometric resolution the precise geometry of a strain pattern in order to correlate it with the magnitude of the corresponding pseudo-magnetic field (see for instance Ref. \cite{Peeters2013-2} for graphene under triaxial stress). Theoretical models to represent nanobubbles in graphene are mainly based on
a gaussian approximation for the strain field, that leads to a non-uniform pseudo-magnetic field possessing a well defined compact support in the spatial domain \cite{Peeters2013,Arias_015}. On the other hand, experimental
STEM measurements \cite{Levy_10,Klimov_012, Yan_012} are consistent with a nearly uniform pseudomagnetic field over a circular
region with a radius commensurate to the size of the nanobubble \cite{Klimov_012} (15 - 25 nm typically), or the width of a ridge \cite{Yan_012}. Ab-initio calculations support these experimental findings as well \cite{Zhu_015}. Interesting perspectives to use this effect
in nanoscale devices have been discussed, for instance by the construction of strain superlattices \cite{Pellegrino2012,Zhang_017}. 

On the other hand, electronic conductance is relatively straightforward to measure, and here we show that it can be directly
correlated with the magnitude of the strain field imposed, thus providing a proof-of-principle for
the development of a piezoelectric sensor with nanometric resolution. In what follows, we shall present a theoretical model to represent elastic scattering of conduction electrons through a graphene nano bubble, represented as a 
disk-shaped region submitted to mechanical strain and an external magnetic field normal
to the plane, as depicted in Fig. \ref{fig1}. We shall obtain exact analytical solutions for the eigenstates within the region, as 
well as for the states scattered off the region. By calculating the differential and total scattering cross-sections, we obtain the transmission coefficient \cite{Tudorovskiy_12,Peres_10} through the region, and calculate the Landauer conductance \cite{Peres_10} for a given bias applied. Our analytical results show explicitly how a combination of a physical magnetic field and mechanical strain leads to
valley-polarization and filtering of the current \cite{Low_010, Settnes_016}.

\section{Model} 
\label{sec:model}
Let us start by writing the effective Dirac Hamiltonian for graphene, involving both valleys $\mathbf{K}_{\pm}=\pm\frac{4\pi}{3\sqrt{3 }a}\hat{\mathbf{e}}_x$, in the presence of generalized SU(2) gauge fields \cite{Gopalakrishnan2012}
\begin{equation}
\hat{H} = \hbar v_{F}\left[
\Gamma_{x}\left( 
\hat{p}_{x} - \sum_{i=1}^{3}A_{x}^{i}Q_{i}
\right) + \Gamma_{y}\left(
\hat{p}_{y} - \sum_{i=1}^{3}A_{y}^{i}Q_{i}
\right)
\right].
\label{eq1}
\end{equation}

Here the Fermi velocity $v_F \sim c/300\sim 10^6$ m/s. We have defined the matrices
\begin{eqnarray}
\Gamma_{x} &=& \hat{\tau}_{3}\otimes\hat{\sigma}_{1},\,\,\,\,\Gamma_{y} = \hat{\tau}_{0}\otimes\hat{\sigma}_{2},\nonumber\\
Q_{0} &=& \hat{\tau}_{0}\otimes\hat{\sigma}_{0},\,\,\,\,Q_{1} = -\hat{\tau}_{2}\otimes\hat{\sigma}_{2},
\nonumber\\
Q_{2} &=& \hat{\tau}_{1}\otimes\hat{\sigma}_{2},\,\,\,\,Q_{3} = \hat{\tau}_{3}\otimes\hat{\sigma}_{0},
\label{eq2}
\end{eqnarray}
where $\sigma$ and $\tau$ are the Pauli matrices acting on the sublattice and valley spaces respectively. The spinor structure over which the Hamiltonian operates is
\begin{eqnarray}
\Psi = \left(
\begin{array}{c}
\psi_{A}^{+}\\ \psi_{B}^{+} \\ \psi_{A}^{-} \\ \psi_{B}^{-}
\end{array}
\right) \equiv \left(
\begin{array}{c}
\Psi^{(+)}\\ \Psi^{(-)}
\end{array}
\right).
\label{eq13}
\end{eqnarray}

To introduce the effect of an external magnetic field and mechanical strain, we chose the following gauge fields:
\begin{eqnarray}
A_{x}^{0} &=& -y\frac{B_{0}}{2}, \,\,\,
A_{y}^{0} = x\frac{B_{0}}{2},\,\,\,
A_{x}^{3} = -y\frac{B_{S}}{2},\,\,\,A_{y}^{3} = x\frac{B_{S}}{2},\nonumber\\
A_{x}^{1} &=& A_{y}^{1} = A_{x}^{2} = A_{y}^{2}=0,
 \label{eq3}
\end{eqnarray}
where $B_{0}$ represents the magnitude of the physical, external magnetic field, while $B_{S}$ characterizes the magnitude of the pseudo-magnetic field induced by mechanical strain. 
The $\mathbf{A}^1$ and $\mathbf{A}^2$ gauge fields, that may be used to model charge density waves \cite{Gopalakrishnan2012}, are set to zero since these phenomena are not under consideration in our present analysis.
For notational convenience, let us define the combination
\begin{eqnarray}
B_{\xi} = B_{0} +\xi B_{S},
\label{eq4}
\end{eqnarray}
representing the effective magnetic field acting at each of the two non-equivalent Dirac cones centered at the wave-vectors $\bm{K}_{\xi} = \xi\frac{4\pi}{3\sqrt{3}a}\hat{\mathbf{e}}_{x}$, for $\xi=\pm$ respectively.

In this case, the Hamiltonian in Eq. \eqref{eq1} has the block diagonal form:

\begin{eqnarray}
\hat{H} = \left(
\begin{array}{cc}
\hat{H}^{+} & 0\\0 & \hat{H}^{-}
\end{array}
\right),
\label{eq6}
\end{eqnarray}
where we have defined
\begin{align}
\hat{H}^{\xi} =& \xi\hbar v_{F}\left[\hat{\sigma}_{1}\left( \hat{p}_{x} + \frac{y}{2}B_{\xi} \right)+\xi\hat{\sigma}_{2}\left(
\hat{p}_{y} - \frac{x}{2}B_{\xi}
\right)\right].
\label{eq7}
\end{align}

A more symmetric, and hence more convenient representation of the Hamiltonian in Eq. \eqref{eq6}
is obtained by transforming the spinor in Eq.(\ref{eq13}) as follows
\begin{eqnarray}
\Psi \rightarrow \tilde{\Psi} = \hat{S}\Psi = \left(
\begin{array}{c}
\psi_{A}^{+}\\ \psi_{B}^{+} \\ \psi_{B}^{-} \\ \psi_{A}^{-}
\end{array}
\right), 
\end{eqnarray}
where we have defined the unitary transformation $\hat{S} = \hat{S}^{-1}$ by the matrix
\begin{eqnarray}
\hat{S} = \left(  
\begin{array}{cc}
\hat{\sigma}_{0} & 0\\0 & \hat{\sigma}_{1}
\end{array}
\right),
\end{eqnarray}
where $\hat{\sigma}_0$ is the $2\times 2$ identity matrix.
The transformed Hamiltonian, after Eq. \eqref{eq6}, is given by
\begin{eqnarray}
\hat{H} \rightarrow \hat{H}_S = \hat{S}\hat{H}\hat{S}^{-1} = 
\left(
\begin{array}{cc}
\hat{H}^{+} & 0 \\
0 & \hat{\sigma}_{1}\hat{H}^{-}\hat{\sigma}_{1} \end{array}
\right)
 \equiv  \left(
\begin{array}{cc}
\hat{H}_{S}^{+} & 0\\ 
0 & \hat{H}_{S}^{-}
\end{array}
\right).
\end{eqnarray}

In particular, making use of the identity $\hat{\sigma}_{1}\hat{\sigma}_{2}\hat{\sigma}_{1} = -\hat{\sigma}_{2}$, we have for the diagonal components of the transformed Hamiltonian
\begin{equation}
\hat{H}_{S}^{\xi} =\xi\hbar v_{F}\left[\hat{\sigma}_{1}\left( \hat{p}_{x} + \frac{y}{2}B_{\xi} \right)
+ \hat{\sigma}_{2}\left(
\hat{p}_{y} - \frac{x}{2}B_{\xi}
\right)\right].
\label{eq71}
\end{equation}

In terms of the definitions above, we have the block-diagonal
eigenvalue problem

\begin{equation}
\left(
\begin{array}{cc}
\hat{H}_{S}^{(+)}-E^{(+)} & 0\\
0 & \hat{H}_{S}^{(-)}-E^{(-)}
\end{array}
\right)\left(
\begin{array}{c}
\tilde{\Psi}^{(+)}\\ \tilde{\Psi}^{(-)}
\end{array}
\right) = 0,
\label{eq14}
\end{equation}
that reduces to two independent eigenvalue problems for the block Hamiltonians $\hat{H}_{S}^{\pm}$
at each valley $\mathbf{K}_{\pm}$.

The eigenvalue problem was solved analytically, with further technical details presented in Appendix \ref{appendixA}. In particular, the energy eigenvalues for the extended Hamiltonian describing both cones under the presence of magnetic
and strain fields are found to be
\begin{align}
E_{\lambda}^{\xi}(n) = \lambda\hbar v_F \sqrt{2 n |B_{\xi}|},\,\,\,\,n = 0, 1, \ldots
\label{eq_Energy}
\end{align}
with $\lambda = \pm 1$ representing particle (hole) eigenstates, while $\xi = \pm$ being the valley index. 
The spinor eigenvectors, for $n>0$, are given by 
\begin{align}
&\tilde{\Psi}_{n,m}^{\xi,\lambda}(r,\phi) = C_{m,n}^{\xi,\lambda}\left(\begin{array}{cc}z^{\frac{|m|}{2}} e^{-z/2} L_{n_{\rho}}^{|m|}(z)e^{i m\phi}\\i\,\alpha_{n}^{\xi}z^{\frac{|m+1|}{2}}e^{-z/2} L_{n'_{\rho}}^{|m+1|}(z) e^{i (m+1)\phi} \end{array}\right)
\label{eq_eigenvector}
\end{align}
where we have defined the dimensionless variable $z = |B_{\xi}|r^2/2$, and $L_n^m(z)$ are the associated Laguerre polynomials \cite{Gradshteyn}.
The coefficients in Eq. \eqref{eq_eigenvector} are defined, for $n>0$ by:
\begin{eqnarray}
n_{\rho} &=& n - \theta(-B_{\xi}) - \frac{|m| - m\,{\rm{\sgn}}B_{\xi}}{2}\nonumber\\
n'_{\rho} &=&  n_{\rho} - \theta(B_{\xi}) + \theta(-m)\nonumber\\
\alpha_{n}^{\xi} &=&\lambda\xi  n^{\theta(-m) - 1/2},
\label{eq_param}
\end{eqnarray}
with $\theta(x)$ the Heaviside step function, and the normalization cefficients
\begin{align}
C_{m,n}^{\xi,\lambda} =& \left( \frac{|B_{\xi}|}{2\pi} \right)^{\frac{1}{2}}
\left\{ 
\frac{\Gamma(|m| + n_{\rho} + 1)}{n_{\rho}!}+ \left(\alpha_n^{\xi} \right)^2\frac{\Gamma(|m+1| + n'_{\rho} + 1)}{n_{\rho}^{'}!}
 \right\}^{-1/2}.
\label{eq_norm}
\end{align}
The index $m$ is an integer, and $\Gamma(z)$ represents the Gamma function. For $\sgn B_{\xi} = +1$, we have $-n \le m < +\infty$, while for $\sgn B_{\xi} = -1$ we have $-\infty < m \le n-1$.

The state with $n = 0$ is given, for $\sgn B_{\xi} = +1$, with $m \ge 0$ by the expression
\begin{equation}
\tilde{\Psi}_{0,m\ge0}^{\xi,\lambda}(r,\phi) = C_{m\ge0,0}^{\xi,\lambda}\left(\begin{array}{c} z^{\frac{m}{2}} e^{-z/2} e^{i m \phi}\\ 0 \end{array} \right).
\end{equation}
On the other hand, for $\sgn B_{\xi} = -1$, the state $n=0$ with $m < 0$ is given by
\begin{equation}
\tilde{\Psi}_{0,m<0}^{\xi,\lambda}(r,\phi) = C_{m<0,0}^{\xi,\lambda}\left(\begin{array}{c}0\\ z^{\frac{|m+1|}{2}} e^{-z/2} e^{i (m+1) \phi} \end{array} \right).
\end{equation}
Here, the normalization coefficients are given by
\begin{align}
C_{m,0}^{\xi,\lambda} =&\left( \frac{|B_{\xi}|}{2\pi}\right)^{1/2}\left\{
\theta (B_{\xi}) \Gamma(|m| + 1)+ \theta(-B_{\xi})\Gamma(|m+1| + 1)
\right\}^{-1/2}.
\label{eq_C0}
\end{align}

\section{Scattering through a nanobuble with magnetic field and mechanical strain}
\label{sec:scattering}

Let us now consider the problem of transport through a graphene sheet submitted to a physical magnetic field and an induced pseudo-magnetic field due to mechanical strain. 
Experimentally, STEM measurements \cite{Levy_10,Klimov_012, Yan_012} reveal that when graphene is submitted to local strain patterns,
the resulting pseudomagnetic fields possess a well defined compact support in the spatial domain. A number of attempts have been published in the literature to model such patterns by a gaussian distributed field, and the corresponding models for the associated Dirac single-particle eigenstates and energy eigenvalues can only be studied numerically \cite{Peeters2013}. More recently, Bahamon \etal \cite{Bahamon_015} studied the conductance induced by different strain nanobubles numerically using molecular dynamics and tight-binding simulations. However, STEM experiments reveal that the magnitude of the pseudomagnetic field due to local strain patterns
is nearly uniform within a region with a characteristic radius
on the order of $15 - 25$ nm \cite{Levy_10,Klimov_012, Yan_012,Zhu_015}. 

Based on the previous statements, we prefer to study the system within a realistic approximation that allows us to obtain analytical solutions. We thus assume that the fields are non-zero only within a circular region of radius $a \sim 15 - 25$ nm (see Fig.\ref{fig1}). We consider then the problem of two-dimensional elastic scattering of an incident free spinor with momentum $\mathbf{k} = (k,0)$ and energy $E_{k,\lambda}^{\xi} = \lambda \hbar v_F |\mathbf{k}|$ ($v_F \sim 10^{6}$ m/s), with $\lambda = \pm 1$ the ``band'' index, and $\xi = \pm 1$ referring to each valley $\mathbf{K}_{\xi}$, respectively.  We will give below a detailed description that generalizes the method in Ref.\cite{adhikari1986} to the case of Dirac fermions. Using this method we will be able to compute the differential scattering cross-section that will be used in the next section to compute the electronic transport.

We begin by considering a free spinor eigenstate incident from the left towards the circular scattering center. This spinor is given by the solution of Eq. \eqref{eq71} with $B_\xi=0$, and thus is given by:
\begin{equation}
\tilde{\Psi}_{in}^{(\lambda,\xi)}(r,\phi) = \frac{1}{\sqrt{2}}\left(\begin{array}{c}1 \\ \lambda\xi \end{array} \right) e^{i k r\cos\phi}.
\label{eq:incm}
\end{equation}
We now proceed with the standard partial wave analysis for scattering. Let us first consider the general solution for the problem in the absence
of external fields and interactions. Following the procedure described in detail in Appendix \ref{appendixA}, we have that the spinor corresponding to the eigenvalue $\hbar m_j$ of the total angular
momentum operator $\hat{J}_3 = \hat{L}_3 + \hat{\sigma}_3/2$ is of the form
\begin{equation}
\tilde{\Psi}_{m_j}(r,\phi) = r^{-1/2}\left(\begin{array}{c} f_{m_j}(r) e^{i(m_j -1/2)\phi}\\-i\,g_{m_j}(r) e^{i(m_j + 1/2)\phi} \end{array}\right).
\label{eq:free}
\end{equation}
The partial wave decomposition, i.e. the resolution into angular momentum channels $m_j$, of the Dirac equation then reduces to the effective coupled eigenvalue problem for the radial functions
\begin{eqnarray}
\left(\begin{array}{cc} - \lambda\xi k &\hat{D}^{\dagger}\\  \hat{D} & - \lambda \xi k\end{array} \right) \left(\begin{array}{c} f_{m_j}\\g_{m_j}
\end{array} \right) = 0,
\label{Eq_Dirac_rad}
\end{eqnarray}
where we have defined (see Appendix \ref{appendixA}) the differential operators $\hat{D} = \frac{d}{dr} - \frac{m_j}{r}$
and its adjoint $\hat{D}^{\dagger} = -\frac{d}{dr} - \frac{m_j}{r}$.
Solving for the system Eq.(\ref{Eq_Dirac_rad}), the radial components of the spinor are determined
through the diagonal eigenvalue problem 
\begin{equation}
\left( \begin{array}{cc} \hat{D}^{\dagger}\hat{D} - k^2 & 0\\0 & \hat{D} \hat{D}^{\dagger}- k^2 \end{array}\right)
\left( \begin{array}{c} f_{m_j}\\g_{m_j}\end{array}\right) = 0.
\label{eq:ev}
\end{equation}
The matrix Eq. \eqref{eq:ev} leads to the pair of differential equations
\begin{align}
\left[-\frac{d^2}{dr^2} + \frac{(m_j - 1/2)^2}{r^2} - k^2 \right] f_{m_j}(r) &= 0,\nonumber\\
\left[-\frac{d^2}{dr^2} + \frac{(m_j + 1/2)^2}{r^2} - k^2 \right] g_{m_j}(r) &= 0.
\label{eq:radial}
\end{align}
The general solution of the system in Eq.(\ref{eq:radial}) is expressed in terms of Bessel functions of the first and second kind,
\begin{align}
f_{m_j}(r) =& c_1 \sqrt{kr} J_{m_j -1/2}(kr) + c_2 \sqrt{kr} Y_{m_j - 1/2}(kr),\nonumber\\
g_{m_j}(r) =& c_3 \sqrt{kr} J_{m_j+1/2}(kr) + c_4 \sqrt{kr} Y_{m_j+1/2}(kr).
\label{eq:bessel}
\end{align}
However, the Dirac equation in its first-order differential form Eq.(\ref{Eq_Dirac_rad}) imposes
a relation between the upper and lower components, i.e.
\begin{eqnarray}
g_{m_j}(r) &=& \frac{\lambda\xi}{k} \hat{D} f_{m_j}(r)=-\lambda\xi\left[ c_1 \sqrt{kr} J_{m_j +1/2}(kr) + c_2 \sqrt{kr} Y_{m_j + 1/2}(kr)\right]
\label{eq:g_free}
\end{eqnarray}
where we applied the Bessel function identity \cite{Gradshteyn}
\begin{eqnarray}
\frac{dZ_{\alpha}}{dx} - \frac{\alpha}{x} Z_{\alpha}(x) = -Z_{\alpha +1}(x).
\end{eqnarray}
The result in Eq.(\ref{eq:g_free}) clearly fixes $c_3 = -c_1$ and $c_4 = -c_2$ in Eq.(\ref{eq:bessel}).

\subsection{Phase-shift}
In elastic scattering theory, the phase shift captures the effect of a scattering region over the transmitted particle waves. In order to express the phase shift associated to the circular region depicted in Fig.\ref{fig1}, let us first consider the asymptotic properties of the Bessel functions \cite{Gradshteyn}, for $k r \gg 1$,
\begin{eqnarray}
J_m(kr) &&\sim \sqrt{\frac{2}{\pi k r}}\cos\left( kr - \left(m + \frac{1}{2}\right)\frac{\pi}{2}\right),\nonumber\\
Y_m(kr) &&\sim \sqrt{\frac{2}{\pi k r}}\sin\left( kr - \left(m + \frac{1}{2}\right)\frac{\pi}{2}\right).
\label{eq:bexp}
\end{eqnarray}
Taking these properties into account, we have that the asymptotic form for the general spinor solution Eq.(\ref{eq:free}) with angular momentum $m \equiv m_j - 1/2$ is, for $k r\gg 1$
\begin{align}
&\tilde{\Psi}_{m_j}^{\lambda,\xi}(r,\phi) \sim \sqrt{\frac{2}{\pi k r}}\tilde{C}_m\left(\begin{array}{c}  e^{i m\phi} \cos\left( kr - \left(m + \frac{1}{2}\right)\frac{\pi}{2}+ \delta_m\right)  \\
i \lambda\xi\, e^{i (m+1)\phi} \cos\left( kr - \left(m + \frac{3}{2}\right)\frac{\pi}{2}+ \delta_{m}\right)
\end{array}\right).
\label{eq:asymp}
\end{align}
Here, we have defined the global coefficients and phase shifts by
\begin{align}
\tilde{C}_{m} =& \sqrt{k}\sqrt{c_1^2 + c_2^2},\nonumber\\
\tan\delta_m =& -c_2/c_1.
\label{eq:shift}
\end{align}

To determine the phase shift $\delta_m$ associated to each angular momentum channel $m$, we have to match each spinor component of the general solution Eq. \eqref{eq:free}, and its first derivative, to the corresponding solution inside the region submitted to the effective magnetic field $B_\xi$, at the boundary $r=a$. In particular, for the upper spinor component, we have the following system of equations:
\begin{align}
c_1 \sqrt{k}J_m(k a) + c_2 \sqrt{k}Y_m (ka) &= C_{m,n}^{\lambda,\xi} \left(\frac{|B_{\xi}|a^2}{2}\right)^{\frac{|m|}{2}} e^{-\frac{|B_{\xi}| a^2}{4}} L_{n_{\rho}}^{|m|}\left(\frac{|B_{\xi}|a^2}{2}\right)\\
c_1\,k^{3/2} J_m^{'}(ka) + c_2\,k^{3/2} Y_m^{'}(ka) &= C_{m,n}^{\lambda,\xi}  \left(\frac{|B_{\xi}|a^2}{2}\right)^{\frac{|m|}{2}}|B_{\xi}|a e^{-\frac{|B_{\xi}| a^2}{4}}
\left\{
\left(\frac{|m|}{|B_{\xi}|a^2} - \frac{1}{2} \right) L_{n_{\rho}}^{|m|}\left(\frac{|B_{\xi}|a^2}{2}\right)\right.\nonumber\\
&\left.\quad- L_{n_{\rho}-1}^{|m|+1}\left(\frac{|B_{\xi}|a^2}{2}\right)
\right\}.
\label{eq:sysa}
\end{align}

An exact analytical solution of this linear system yields a closed expression for the phase shift $\delta_m$,
\begin{widetext}
\begin{equation}
\tan\delta_m = \frac{J_{m+1}(ka) + \displaystyle\frac{J_{m}(ka)}{ka}\left\{|m| - m - \frac{|B_{\xi}|a^2}{2}- \frac{L_{n_{\rho}-1}^{|m|+1}(|B_{\xi}|a^2/2)}{L_{n_{\rho}}^{|m|}(|B_{\xi}|a^2/2)}
\right\}}
{Y_{m+1}(ka) + \displaystyle\frac{Y_{m}(ka)}{ka}\left\{
|m| - m - \frac{|B_{\xi}|a^2}{2}- \frac{L_{n_{\rho}-1}^{|m|+1}(|B_{\xi}|a^2/2)}{L_{n_{\rho}}^{|m|}(|B_{\xi}|a^2/2)}
\right\}}.
\label{eq:phase}
\end{equation}
\end{widetext}

Here, we have made use of the following the mathematical identities \cite{Gradshteyn}
\begin{eqnarray}
\frac{d}{dx} L_n^{|m|}(x) &=& - L_{n-1}^{|m|+1}(x),\nonumber\\
\frac{d}{dx}J_{m}(x) &=& \frac{1}{2}\left( J_{m-1}(x) - J_{m+1}(x)\right),\nonumber\\
\frac{d}{dx}Y_{m}(x) &=& \frac{1}{2}\left( Y_{m-1}(x) - Y_{m+1}(x)\right).
\end{eqnarray}

\subsection{Scattering cross section}
In the region $r\gg a$, the state will be given by a linear combination of the incident and
scattered spinor
\begin{equation}
\tilde{\Psi}_{out}(r,\phi) \sim \frac{1}{\sqrt{2}}\left(\begin{array}{c}1 \\ \lambda\xi \end{array} \right) e^{i k r\cos\phi} + \left( \begin{array}{c}f_{1}(\phi)\\f_{2}(\phi)\end{array} \right)\frac{e^{ikr}}{\sqrt{r}},
\label{eq:out}
\end{equation}
with amplitudes $f_1(\phi)$ and $f_2(\phi)$ for each component of the scattered spinor.
 In the same region, we have that this expression must be equal to the asymptotic form of the solution, represented in terms of
phase shifts, Eq.\eqref{eq:asymp}. In order to analyse the contribution of each partial wave with angular momentum $\hbar m$, we use the mathematical identity
\begin{align}
e^{ikr\cos\phi} &= \sum_{m=-\infty}^{\infty}i^{m} e^{im\phi} J_m(kr)\sim\sqrt{\frac{2}{\pi k r}}\sum_m i^m e^{im\phi}  \cos\left(k r - \frac{\pi}{2}\left(m + \frac{1}{2} \right) \right)
\end{align}

Therefore, substituting into Eq.(\ref{eq:out}) we find that:
\begin{align}
\tilde{\Psi}_{out}(r,\phi) =& \sum_m \left(\begin{array}{c} \frac{i^m}{\sqrt{\pi k r}} e^{im\phi}\cos\left( kr - \frac{\pi}{2}\left( m + \frac{1}{2}\right) \right) \\
\lambda\xi\frac{i^{m+1}}{\sqrt{\pi k r}} e^{i{(m+1)}\phi}\cos\left( kr - \frac{\pi}{2}\left( m + \frac{3}{2}\right) \right)
\end{array}\right)+ \left( \begin{array}{c}f_{1}(\phi)\\f_{2}(\phi)\end{array} \right)\frac{e^{ikr}}{\sqrt{r}}.
\label{eq:out3}
\end{align}

Equating expressions (\ref{eq:asymp}) and (\ref{eq:out3}), we demand for the pre-factors of $e^{\pm ikr}$ to be the same
on both sides, thus yielding the following system of equations 

\begin{align}
\sum_{m}\left(\begin{array}{c}\frac{i^m}{2\sqrt{\pi k}} e^{im\phi} e^{-i\pi(m+1/2)/2}\\
\lambda\xi \frac{i^{(m+1)}}{2\sqrt{\pi k}} e^{i(m+1)\phi} e^{-i\pi(m+3/2)/2}
\end{array} \right)+ \left(\begin{array}{c}f_1(\phi)\\f_2(\phi)\end{array}\right)&= \frac{1}{\sqrt{2\pi k}}\sum_m\tilde{C}_m\left(\begin{array}{c} e^{i m\phi} e^{-i\pi(m+1/2)/2 + i\delta_m}\\
\lambda\xi\,i\, e^{i (m+1)\phi} e^{-i\pi(m+3/2)/2 + i\delta_{m}}\end{array}\right)\nonumber\\
\sum_m \left(\begin{array}{c}
\frac{i^{m}}{2\sqrt{\pi k}} e^{i m\phi} e^{i\pi(m + 1/2)/2}\\
\lambda\xi \frac{i^{m+1}}{2\sqrt{\pi k}} e^{i (m+1)\phi} e^{i\pi(m + 3/2)/2}
\end{array}
\right)&= \sum_m \frac{\tilde{C}_m}{\sqrt{2\pi k}} \left(\begin{array}{c}
 e^{i m\phi} e^{i\pi(m+1/2)/2} e^{-i\delta_m}\\
\lambda\xi\,i\, e^{i (m+1)\phi} e^{i\pi(m+3/2)/2} e^{-i\delta_{m}}
\end{array}
\right).
\label{eq:sys1}
\end{align}
Using orthogonality of the basis $\left\{e^{i m \phi}\right\}$, we can compute $\tilde{C}_m$, which is thus given by:
\begin{align}
\tilde{C}_m =& \frac{i^m}{\sqrt{2}} e^{i\delta_m}.
\end{align}

Inserting the previous result into the system of equations Eq.(\ref{eq:sys1}), we solve for the scattering amplitudes:
\begin{equation}
\left( \begin{array}{c} f_1(\phi)\\f_2(\phi)\end{array}\right) = \frac{e^{-i\pi/4}}{2\sqrt{\pi k}}\sum_m\left(\begin{array}{c} e^{im\phi} \\
\lambda\xi e^{i(m+1)\phi}\end{array} \right)\left( e^{2 i \delta_m} -1 \right).
\label{eq:amp}
\end{equation}
The differential scattering cross-section is given by the modulus of the vector above,
\begin{equation}
\frac{d\sigma}{d\phi} = |f_1(\phi)|^2 + |f_2(\phi)|^2,
\label{eq:cross_section}
\end{equation}
and the total scattering cross section (with dimensions of length instead of area) is then given by integrating over the scattering angle $\phi$ ($0\le \phi \le 2\pi$)
\begin{align}
\sigma =& \int_0^{2\pi} \left(  |f_1(\phi)|^2 + |f_2(\phi)|^2  \right)d\phi\nonumber\\
=& \frac{4}{k}\sum_{m=-\infty}^{\infty}\sin^2\delta_{m}.
\label{eq_total_scatt}
\end{align}

\section{Transmission and Landauer ballistic current}

\begin{figure}[tbp]
\centering
\includegraphics[width=0.5\columnwidth]{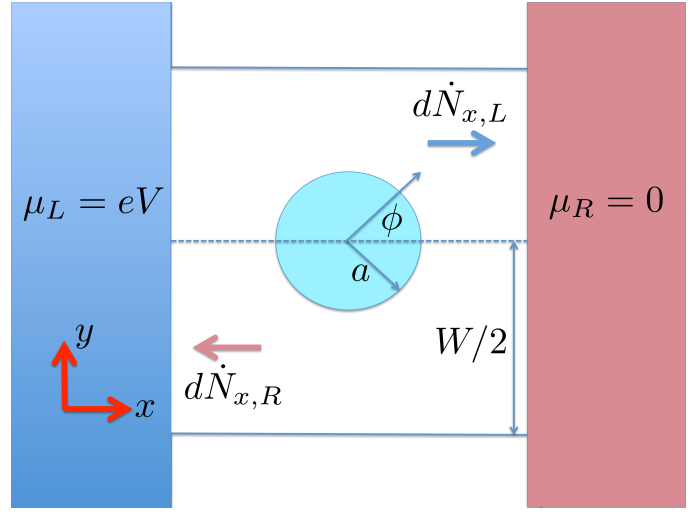}
\caption{(Color online) Pictorial representation of the system. The left and right contacts are assumed as semi-infinite
graphene regions, held at chemical potentials $\mu_{L}$ and $\mu_R$, respectively. The nano bubble is represented as a disk of radius $a$, submitted to the influence of the magnetic field and the mechanical strain.}
\label{fig1}
\end{figure}

Let us consider a graphene stripe, of width $W$ ($y$-direction), which is connected to two semi-infinite graphene contacts
held at chemical potentials $\mu_L$ and $\mu_R$, respectively. As a model for a nanobubble, we shall assume that a circular region
of radius $a$ is submitted to a perpendicular uniform magnetic field ($\hat{\mathbf{e}}_3\,B_0$) and to mechanical
strain as well. Typical experimental values for the characteristic diameter of graphene bubbles are $a \sim 15 - 25 nm$ \cite{Levy_10,Klimov_012, Yan_012,Zhu_015}, while
a graphene ribbon will have typical widths $W \sim 10 \mu m$ . Therefore, under realistic experimental conditions $a/W \ll 1$, and hence any influence of the edges of the ribbon over the carrier dynamics at the nanobubble
becomes negligible. Within the Landauer ballistic picture, the net current along the stripe ($x$-direction) is given by the net counterflow of
the particle currents emitted from the left and right semi-infinite graphene contacts, respectively. Each contact is assumed to be in thermal equilibrium, with the Fermi-Dirac distributions $f(E-\mu_L,T)\equiv f_L(E)$
and  $f_R(E) \equiv f(E-\mu_R,T)$, respectively.  A pictorial description
of the system is shown in Fig. \ref{fig1}.

The particle flux (per unit width) emitted by the left (L) and right (R) contacts, respectively, is defined as
\begin{equation}
dJ_{x,L/R} = v_x D_{L/R}(E) f_{L/R}(E) dE,
\end{equation}
where $D_{L/R}(E)$ is the (surface-normalized) density of states at each contact.

The effect of the nanobubble over charge transport can be expressed as an effective  one-dimensional cross-section $W T_{\xi}(E,\phi)$, with $T_{\xi}(E,\phi)$ the transmission coefficient in the direction specified by the angle $\phi$,
for an incident spinor arising from the valley $\mathbf{K}_{\xi}$. We thus define the effective cross-section in the $\phi$-direction by the expression
\begin{align}
W\, T_{\xi}(E_k,\phi) &= \frac{1}{\sigma(k)} \frac{d\sigma(k)}{d\phi} \sum_{n,\lambda} \delta\left(\lambda\,k - \frac{E_{n}^{\xi}}{\hbar v_F}\right)=  \frac{\hbar v_F}{\sigma} \frac{d\sigma}{d\phi} \sum_{n,\lambda,\xi} \delta\left(E_{k,\lambda} - E_n^{\xi}\right)
\nonumber\\
&= \frac{2\hbar^2 v_F^2}{\sigma \pi}\sum_{m,m',n,\lambda,\xi}\frac{1}{E_{k,\lambda}}e^{i(m-m')\phi}e^{i(\delta_m - \delta_{m'})}\sin\delta_m \sin\delta_{m'} \delta\left(E_{k,\lambda} - E_n^{\xi}\right).
\label{eq_transm}
\end{align}

Here, the differential scattering cross-section is calculated from Eq.(\ref{eq:cross_section}), while 
the total cross section is obtained in terms of the phase shifts by Eq.(\ref{eq_total_scatt}).
The Dirac delta function enforces the energy conservation condition assumed for elastic scattering.

The particle flow (per unit time) along the $x$-direction emitted by the left (L) 
contact and arising from the $\mathbf{K}_{\xi}$ valley is ($v_x = v_F\cos\phi$)
\begin{eqnarray}
d\dot{N}_{x,L}^{\xi} = W\,T_{\xi}(E,\phi)d\phi\, dJ_{x,L} = v_F\cos(\phi)\,D_L(E)\,f_L(E)dE \times W\,T_{\xi}(E,\phi) d\phi,\nonumber
\end{eqnarray} 
with an analogous expression for the right (R) particle flow.
The net electric current flowing across the region will be $I = I_{+} + I_{-}$,
with the valley-polarized component given by 
\begin{eqnarray}
I_{\xi} = e\int\left( d\dot{N}_{x,L}^{\xi} - d\dot{N}_{x,R}^{\xi} \right)= e v_F W \int_{-\infty}^{\infty} dE \left[ D_L(E)f_L(E) - D_R(E)f_R(E)\right] \bar{T}_{\xi}(E).
\label{eq_current}
\end{eqnarray}
Here,  we have defined the net transmission coefficient for
Dirac spinors at valley $\mathbf{K}_{\xi}$ as the angular average $\bar{T}_{\xi}(E) = \int_{-\pi/2}^{\pi/2}d\phi\,\cos\phi\,T_{\xi}(E,\phi)$, that reduces to the analytical expression
\begin{align}
\bar{T}_{\xi}(E) = \frac{4\hbar^2 v_F^2}{\pi W\, E \sigma}\sum_{n,p,m}\frac{(-1)^{p+1}}{4 p^2-1} e^{i(\delta_{m}-\delta_{m-2p})}\left[\sin(\delta_m)\sin(\delta_{m-2p})\,\delta\left( E - E_n^{\xi} \right)\right].
\label{eq_trans_coeff}
\end{align}
where we used the result
\begin{eqnarray}
\int_{-\pi/2}^{\pi/2} d\phi\cos\phi\,e^{i(m-m')\phi} = \delta_{m-m',2p} \frac{2\,(-1)^{p+1}}{4 p^2 -1},\,\,\,p \in \mathbb{Z}.\nonumber\\
\end{eqnarray}
It is important to remark that the valley-polarized transmission coefficients defined by Eq.(\ref{eq_trans_coeff}) are not c-functions, but distributions (a superposition of Dirac-deltas),
and hence it is not possible to plot them graphically. However, as will be discussed in detail later on in the context of the current-voltage characteristics, the
transmission coefficient corresponding to the $\xi=(-)$ valley defines a denser distribution in energy space. The reason is that, for the effective pseudomagnetic field
$B_{-} = B_0 - B_S$ at this valley, the corresponding energy eigenvalues $E_{n}^{\xi} \sim \sqrt{|B_{\xi}|n}$ constitute a denser set than those of the $\xi=(+)$ valley, assuming for
definiteness $B_0 > 0$ and $B_S > 0$. Thus, as will be verified later, the valley-polarized current components should satisfy $I_{-} > I_{+}$.
Assuming both contacts are identical semi-infinite graphene regions, the density of states are equal, and given by
\begin{eqnarray}
D_L(E) &=& D_R(E) = D(E)
= 4 \int\frac{d^2 k}{(2\pi)^2}\left[\delta\left(E - \hbar v_F k \right) + \delta\left(E + \hbar v_F k \right) \right]= \frac{2}{\pi(\hbar v_F)^2}\left[E\theta(E) + (-E)\theta(-E)\right]\nonumber\\
&=& \frac{2|E|}{\pi(\hbar v_F)^2}\theta(|E|),
\end{eqnarray}
where the factor of $4$ arises from the spin and valley degeneracy at each of the graphene semi-infinite contacts.
With this consideration, the expression for the valley-polarized component of the current $I_{\xi}$ becomes
\begin{align}
I_{\xi} &= e v_F W \int_{-\infty}^{\infty} dE D(E) \bar{T}_{\xi}(E) \left[ f_L(E) - f_R(E) \right]\nonumber\\
&=\frac{8 e v_F}{\pi^2}\sum_{n,m,p}\frac{(-1)^{p+1}}{\sigma(E_n^{\xi})(4p^2-1)}
e^{i(\delta_m - \delta_{m-2p})}\sin\delta_m\sin\delta_{m-2p}\left[ f_L(E_n^{\xi}) - f_R(E_n^{\xi})\right],
\label{eq_current}
\end{align}
with the total current given by $I = I_{+} + I_{-}$.
\section{Results and Discussion}

\begin{figure}[hbt]
\centering
\subfigure[]{\includegraphics[width=0.3277\textwidth]{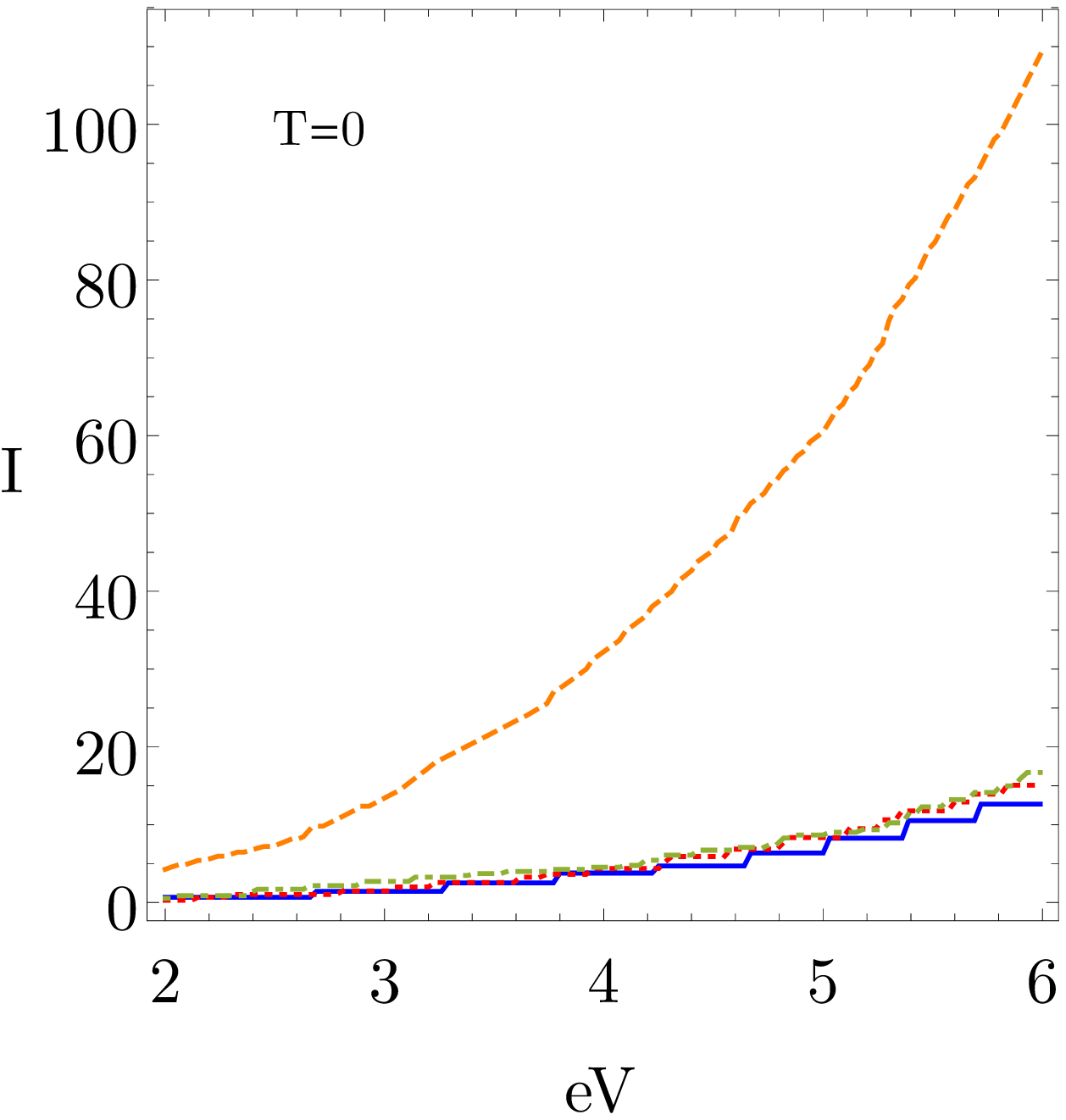}\label{fig2a} }
\subfigure[]{\includegraphics[width=0.3277\textwidth]{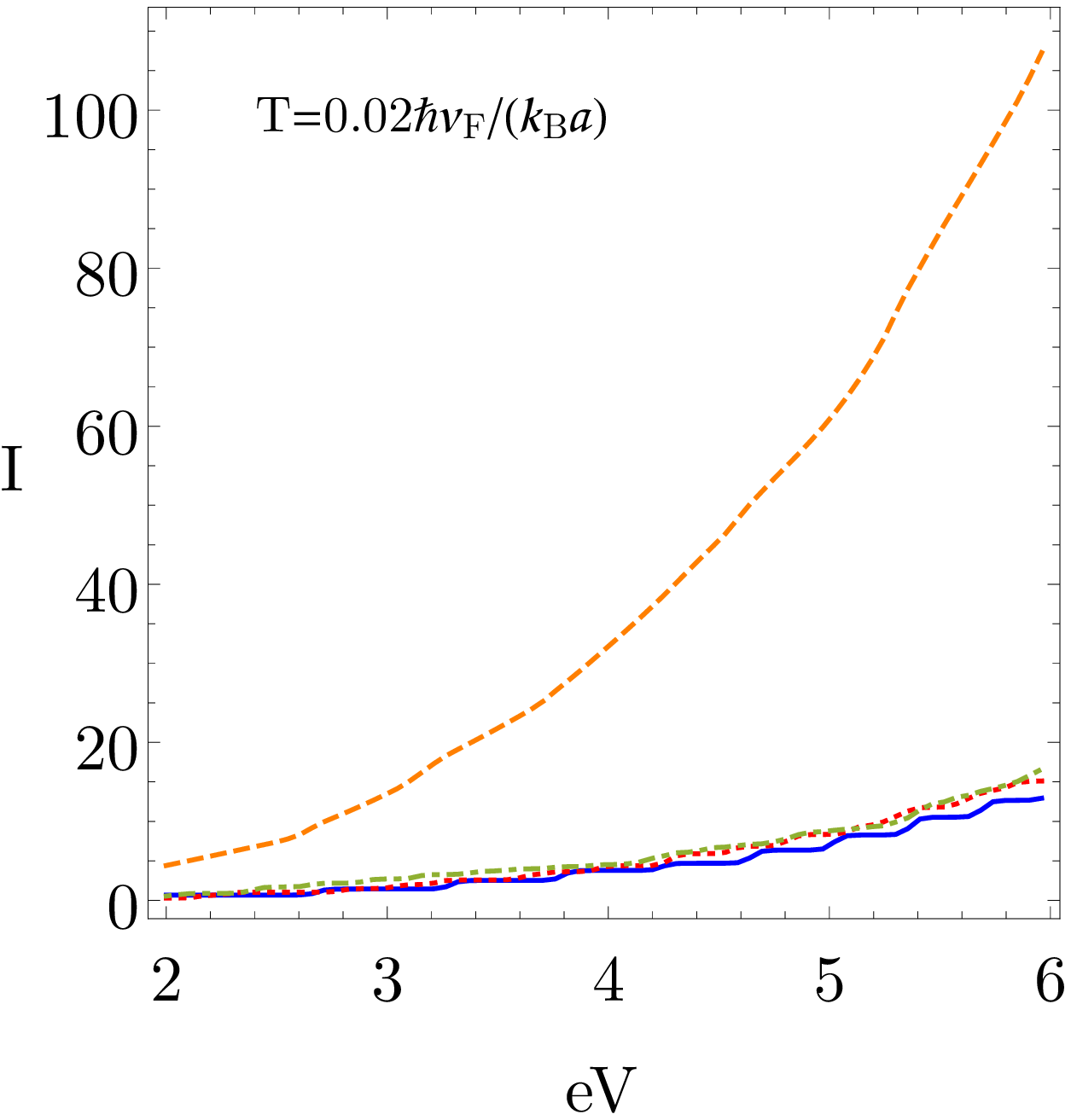}\label{fig2b}}
\subfigure[]{\includegraphics[width=0.3277\textwidth]{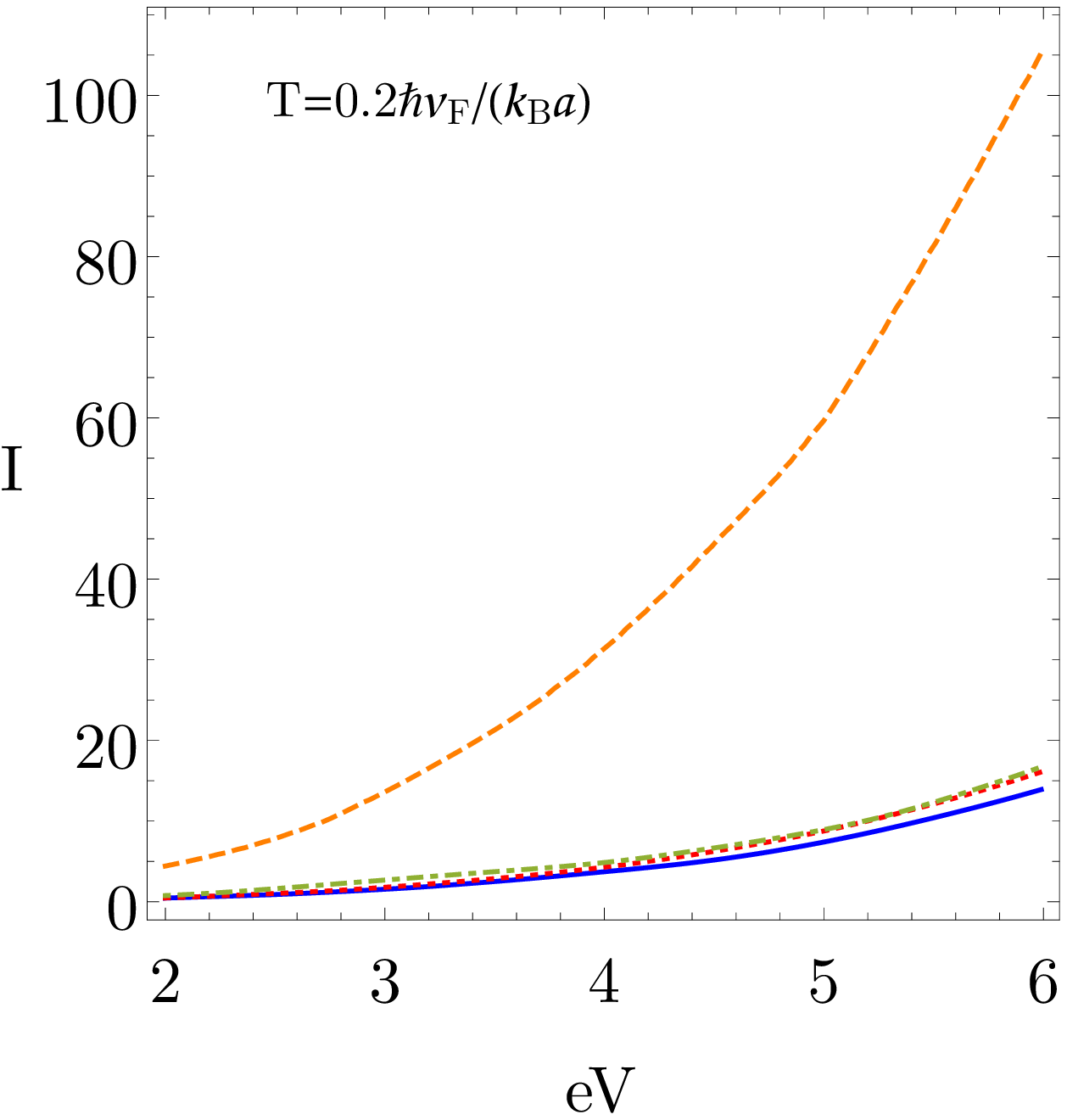}\label{fig2c}}
\caption{(Color online) Current (in units of $e v_F/a$) calculated from the analytical Eq.(\ref{eq_current}), as a function of applied bias $V$ (in units of $\hbar v_F/a$), for fixed $B_0a^2=1.8\tilde{\phi}_0$ and different values of $B_S$. The solid (blue) line corresponds to $B_Sa^2=0$, the dotted (red) line corresponds to $B_Sa^2=0.5\tilde{\phi}_0$, the dotdashed (green) line corresponds to $B_Sa^2=1.1\tilde{\phi}_0$ and the dashed (orange) line corresponds to $B_Sa^2=1.7\tilde{\phi}_0$, with $\tilde{\phi}_{0}\equiv (v_F/c)\hbar /e$. The subfigures (a), (b) and (c) correspond to the different values of the temperature, $T=0$, $T = 0.02 \,\hbar v_F/ (k_B a)$ and $T = 0.2 \,\hbar v_F/ (k_B a)$ respectively.}
\label{fig2}
\end{figure}

\begin{figure}[hbt]
\centering
\subfigure[]{\includegraphics[width=0.3277\textwidth]{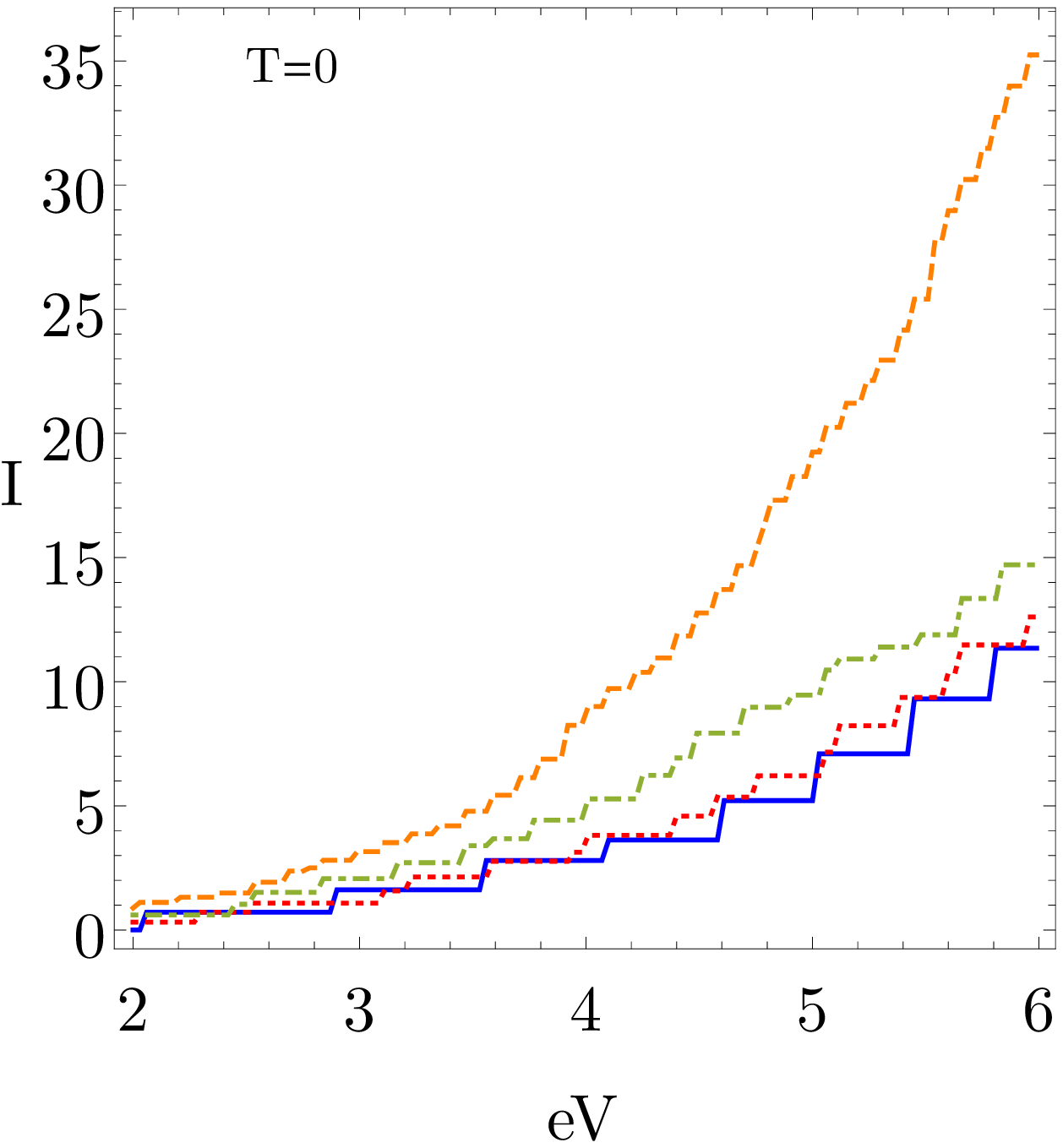}\label{fig3a} }
\subfigure[]{\includegraphics[width=0.3277\textwidth]{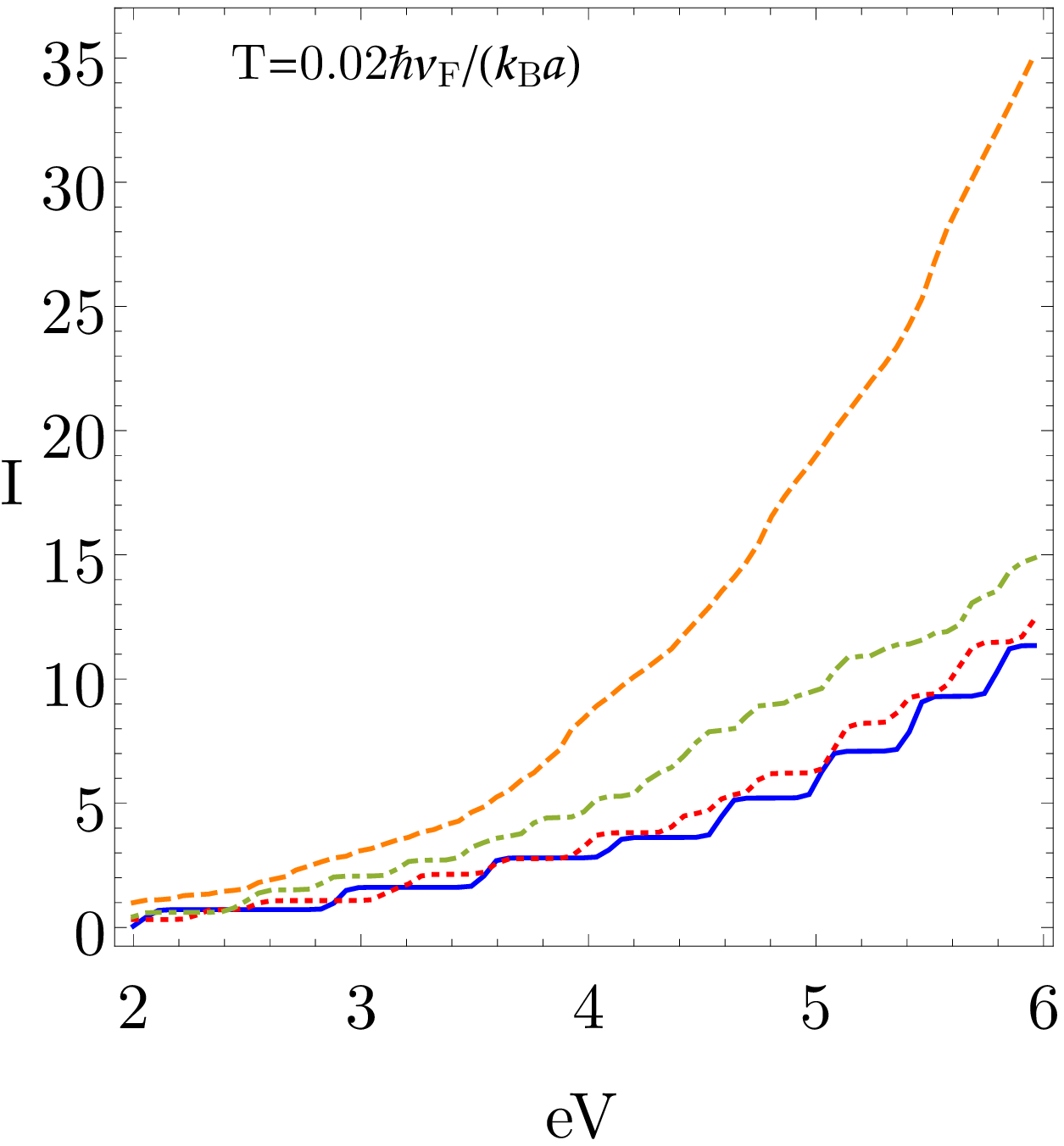}\label{fig3b}}
\subfigure[]{\includegraphics[width=0.3277\textwidth]{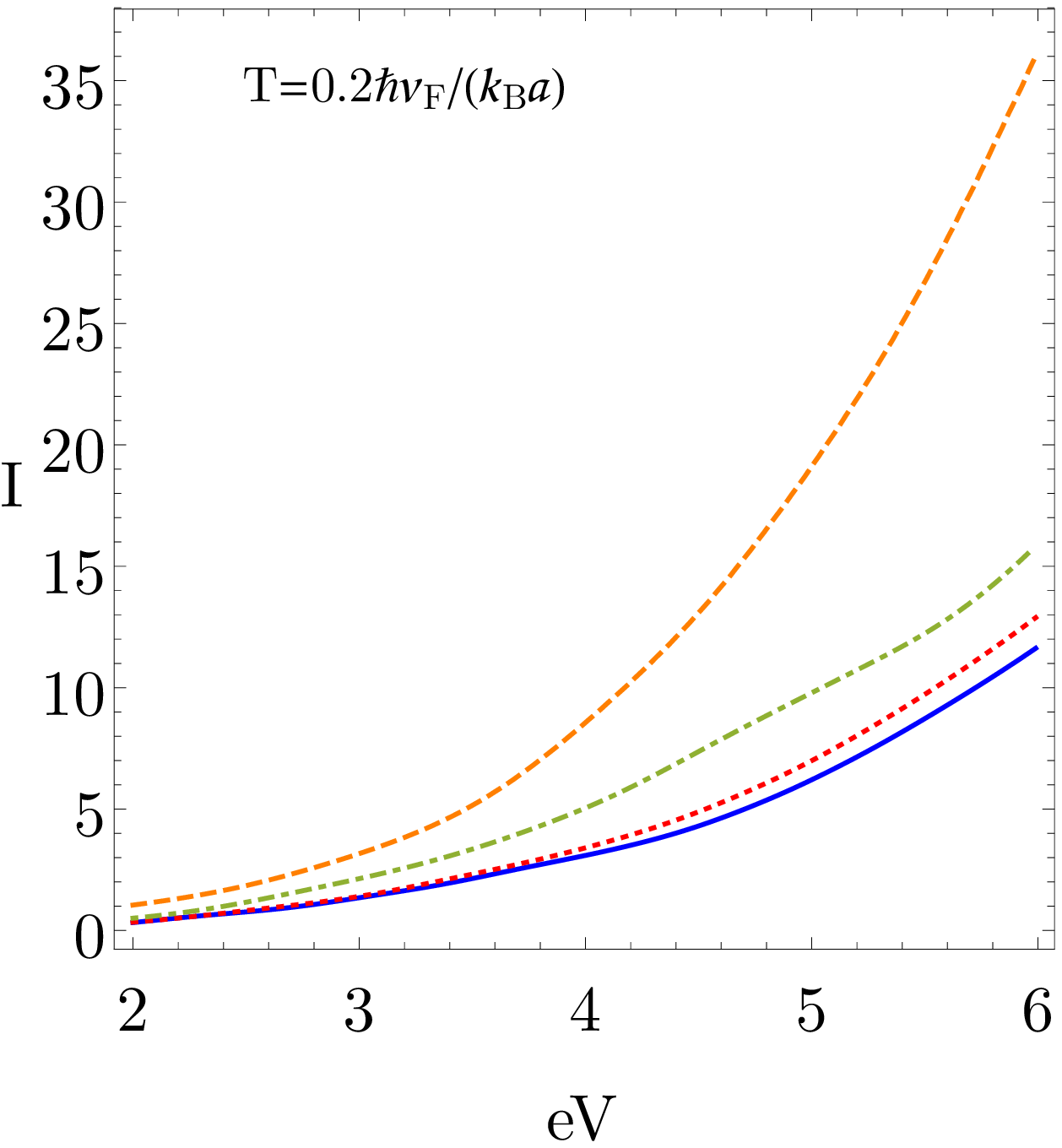}\label{fig3c}}
\caption{(Color online) Current (in units of $e v_F/a$) calculated from the analytical Eq.(\ref{eq_current}) , as a function of applied bias $e V$ (in units of $\hbar v_F/a$), for fixed $B_0a^2=2.1\tilde{\phi}_0$ and different values of $B_S$. The solid (blue) line corresponds to $B_Sa^2=0$, the dotted (red) line corresponds to $B_Sa^2=0.5\tilde{\phi}_0$, the dotdashed (green) line corresponds to $B_Sa^2=1.1\tilde{\phi}_0$ and the dashed (orange) line corresponds to $B_Sa^2=1.7\tilde{\phi}_0$, with $\tilde{\phi}_{0}\equiv (v_F/c)\hbar /e$. The subfigures (a), (b) and (c) correspond to the different values of the temperature, $T=0$, $T = 0.02 \,\hbar v_F/ (k_B a)$ and $T = 0.2 \,\hbar v_F/ (k_B a)$ respectively}
\label{fig3}
\end{figure}

\begin{figure}[hbt]
\centering
\subfigure[]{\includegraphics[width=0.3277\textwidth]{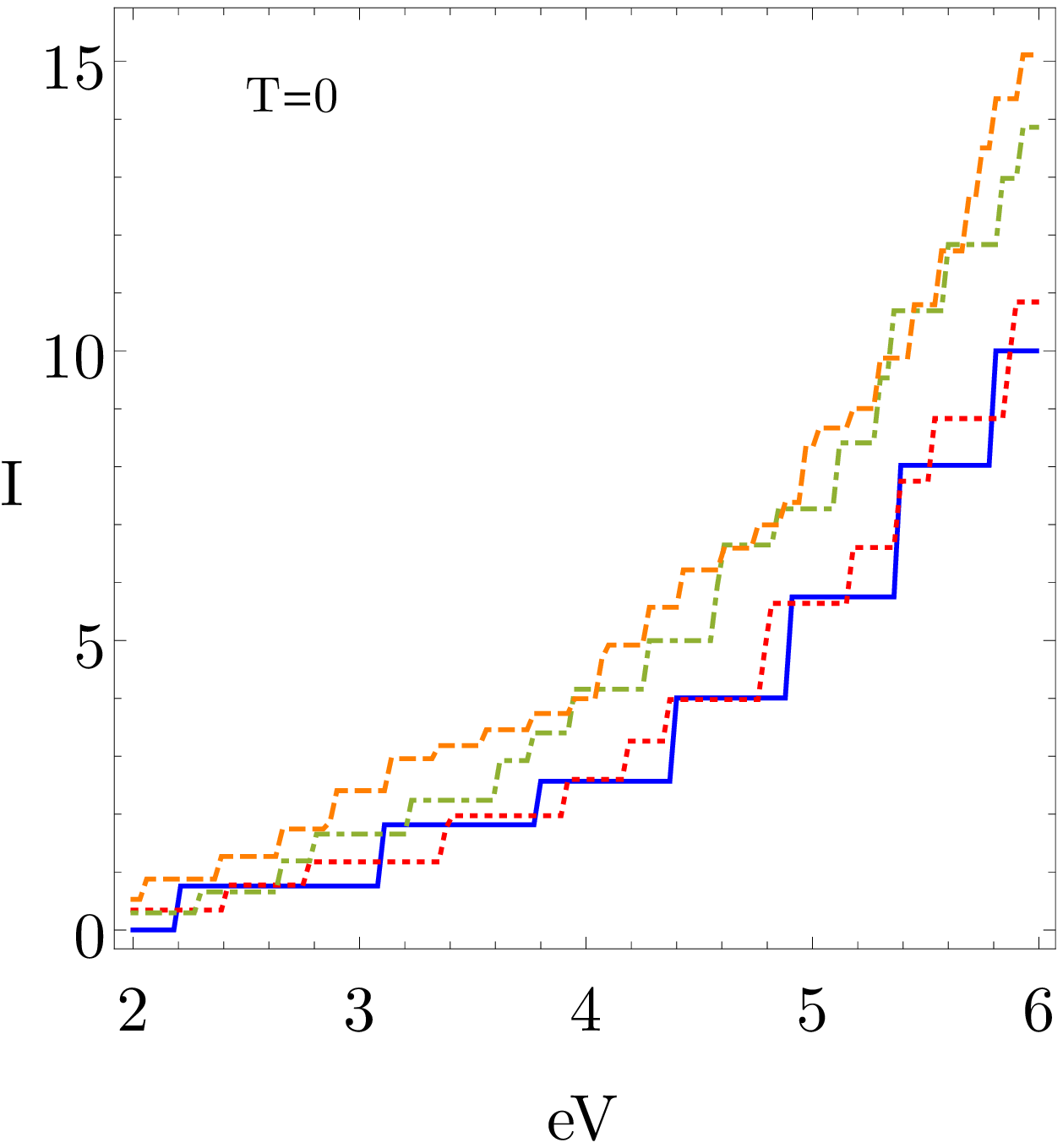}\label{fig4a} }
\subfigure[]{\includegraphics[width=0.3277\textwidth]{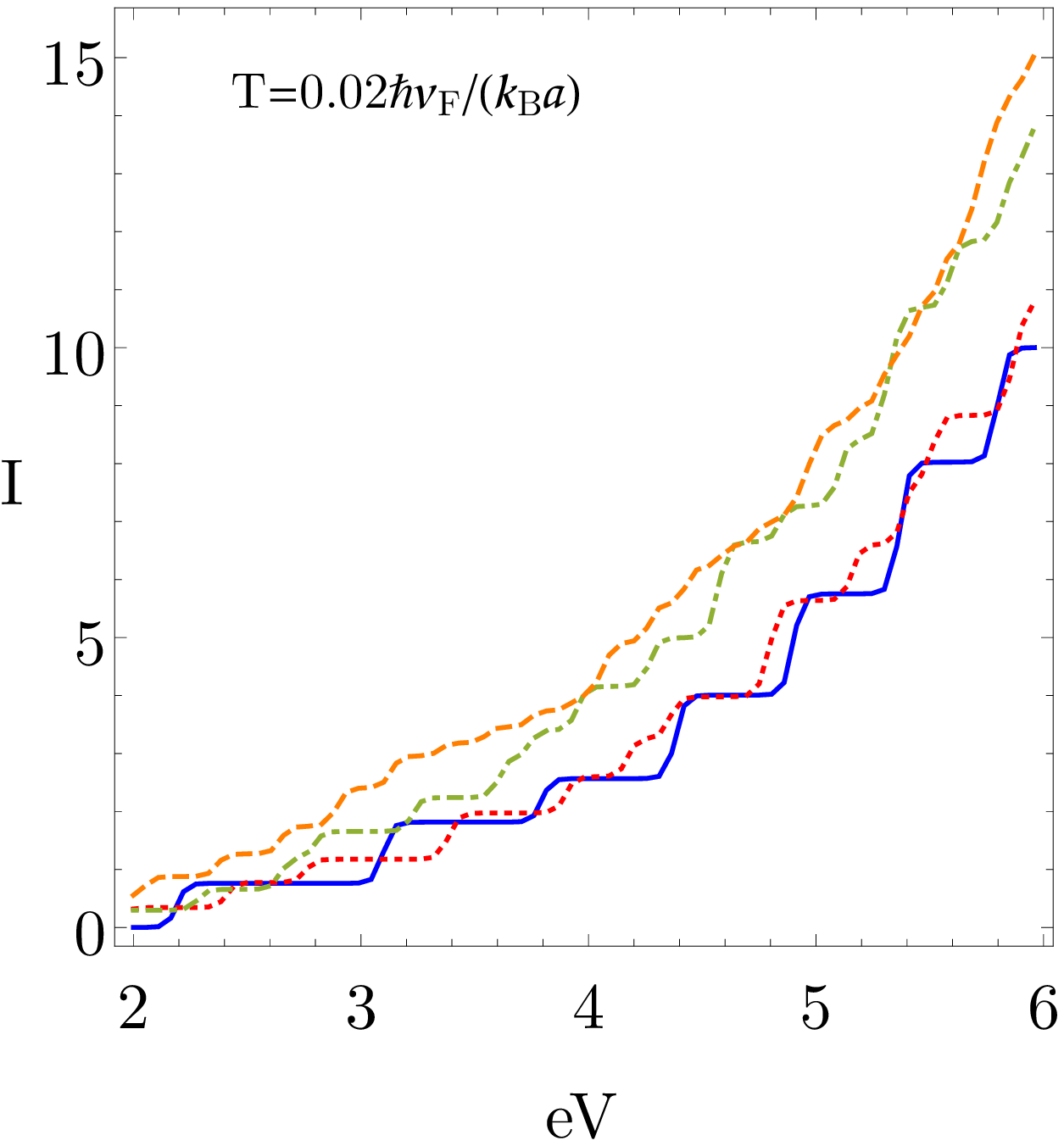}\label{fig4b}}
\subfigure[]{\includegraphics[width=0.3277\textwidth]{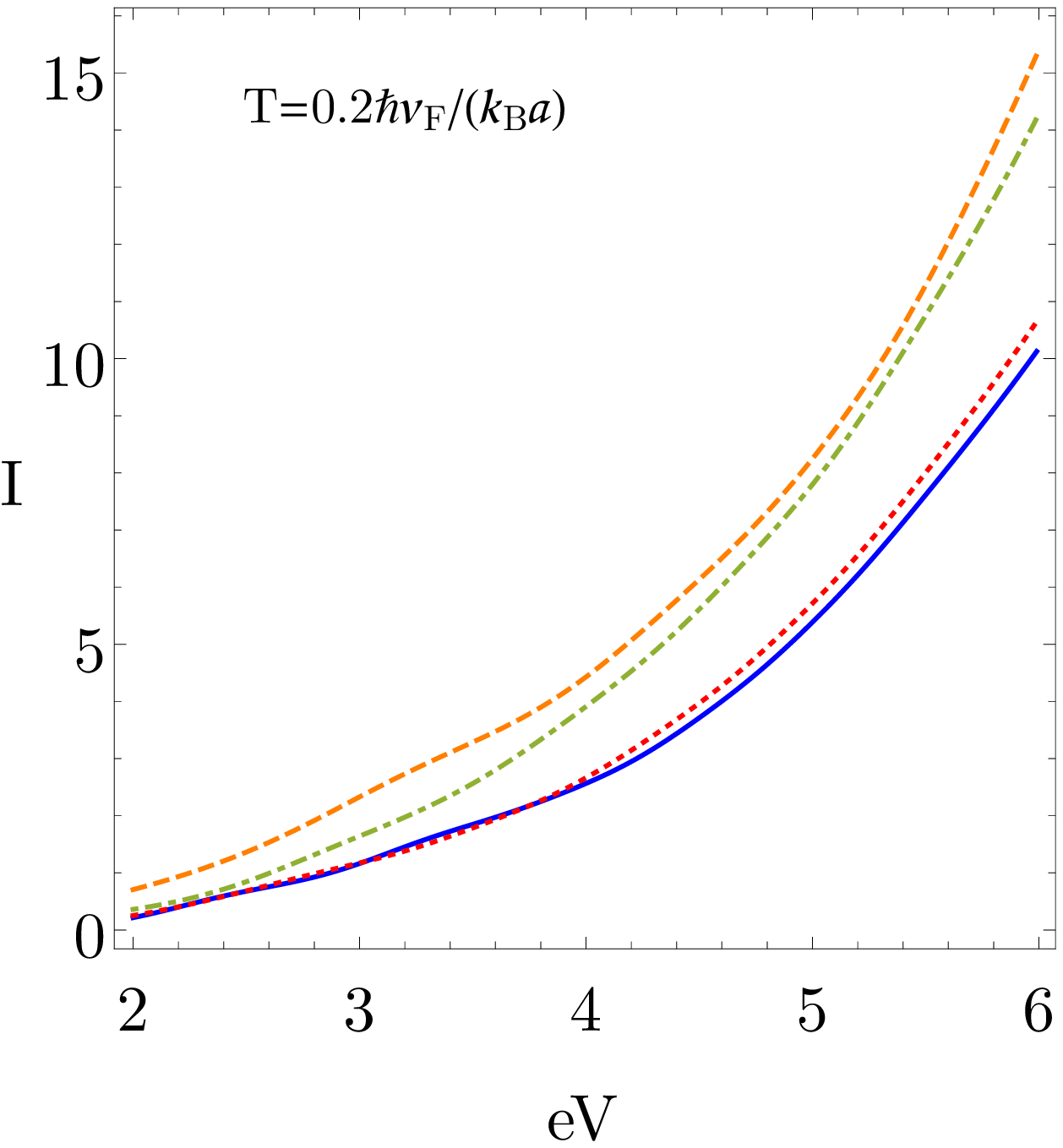}\label{fig4c}}
\caption{(Color online) Current (in units of $e v_F/a$) calculated from the analytical Eq.(\ref{eq_current}), as a function of applied bias $e V$ (in units of $\hbar v_F/a$), for fixed $B_0a^2=2.4\tilde{\phi}_0$ and different values of $B_S$. The solid (blue) line corresponds to $B_Sa^2=0$, the dotted (red) line corresponds to $B_Sa^2=0.5\tilde{\phi}_0$, the dotdashed (green) line corresponds to $B_Sa^2=1.1\tilde{\phi}_0$ and the dashed (orange) line corresponds to $B_Sa^2=1.7\tilde{\phi}_0$, with $\tilde{\phi}_{0}\equiv (v_F/c)\hbar /e$. The subfigures (a), (b) and (c) correspond to the different values of the temperature, $T=0$, $T = 0.02 \,\hbar v_F/ (k_B a)$ and $T = 0.2 \,\hbar v_F/ (k_B a)$ respectively.}
\label{fig4}
\end{figure}

\begin{figure}[hbt]
\centering
\subfigure[]{\includegraphics[width=0.3277\textwidth]{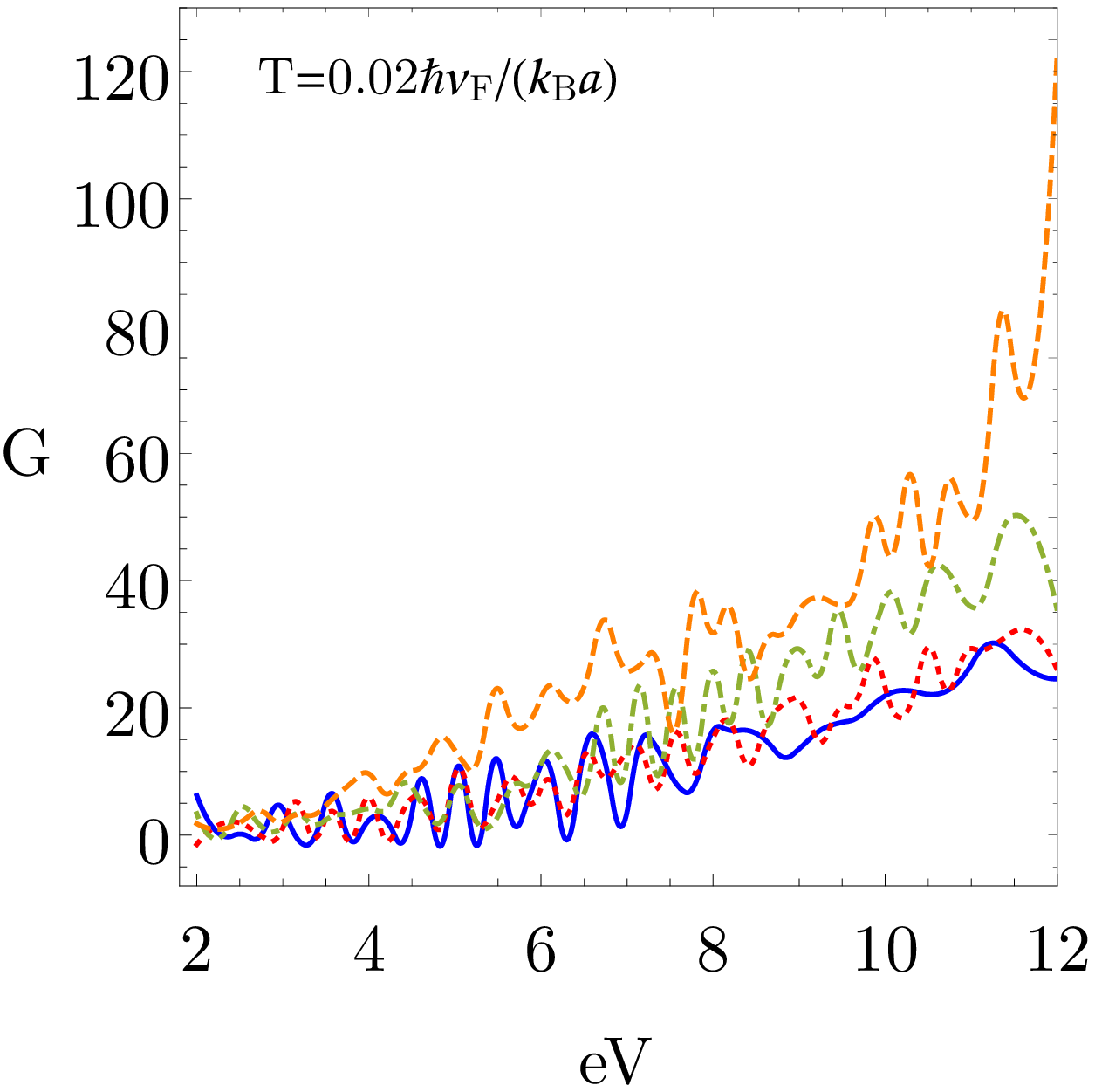}\label{fig5a}}
\subfigure[]{\includegraphics[width=0.3277\textwidth]{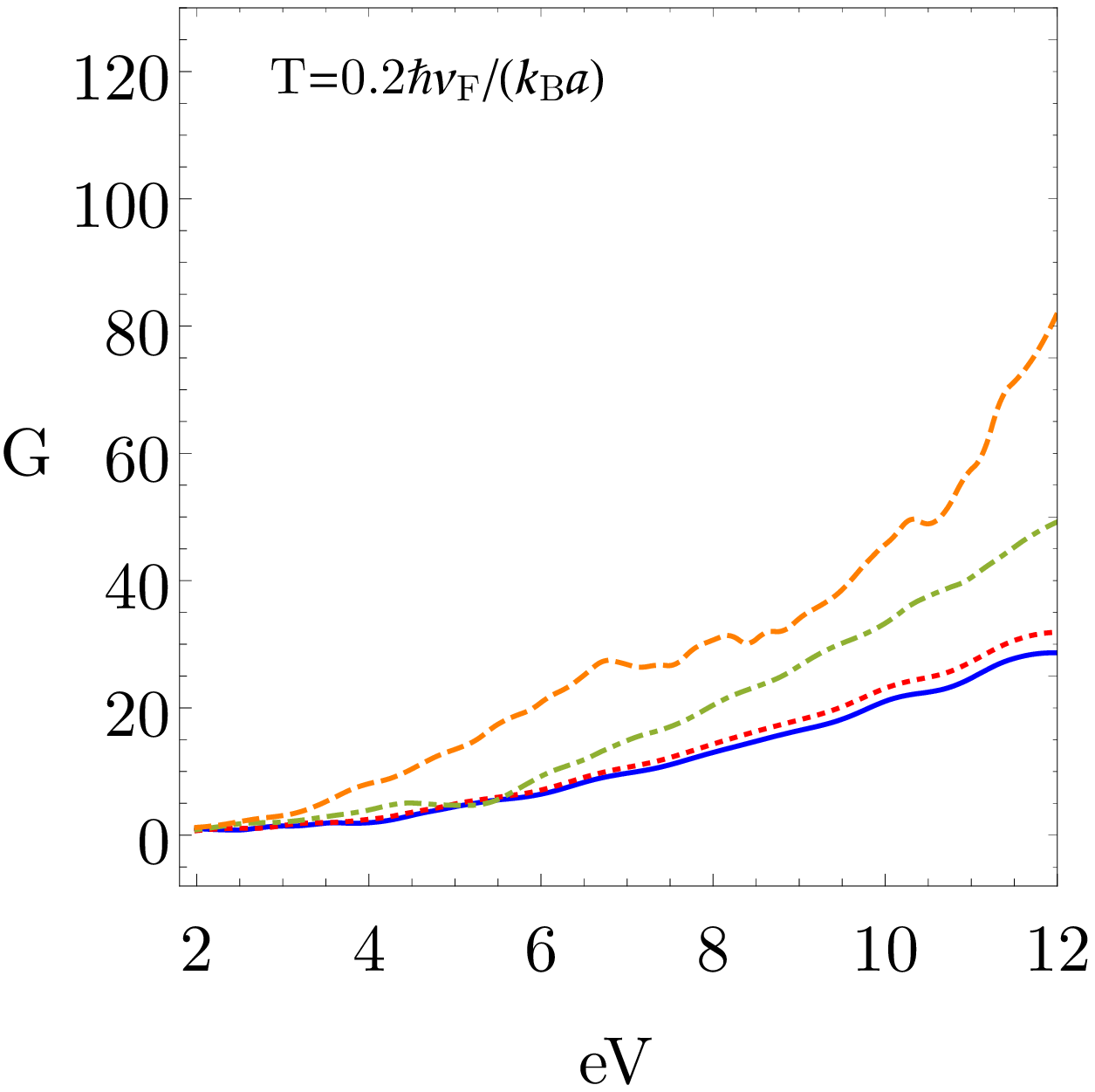}\label{fig5b}}
\caption{(Color online) Conductance (in units of $e^2/\hbar$) as a function of bias $e V$ (in units of $\hbar v_F/a$), calculated as the voltage-derivative of the analytical Eq.(\ref{eq_current}), for fixed $B_0a^2=2.1\tilde{\phi}_0$ and different values of $B_S$.
The solid (blue) line corresponds to $B_Sa^2=0$, the dotted (red) line corresponds to $B_Sa^2=0.5\tilde{\phi}_0$, the dotdashed (green) line corresponds to $B_Sa^2=1.1\tilde{\phi}_0$ and the dashed (orange) line corresponds to $B_Sa^2=1.7\tilde{\phi}_0$, with $\tilde{\phi}_{0}\equiv (v_F/c)\hbar /e$. The subfigures (a) and (b) correspond to the different values of the temperature $T = 0.02 \,\hbar v_F/ (k_B a)$ and $T = 0.2 \,\hbar v_F/ (k_B a)$ respectively.}
\label{fig5}
\end{figure}

\begin{figure}[hbt]
\centering
\subfigure[]{\includegraphics[width=0.3277\textwidth]{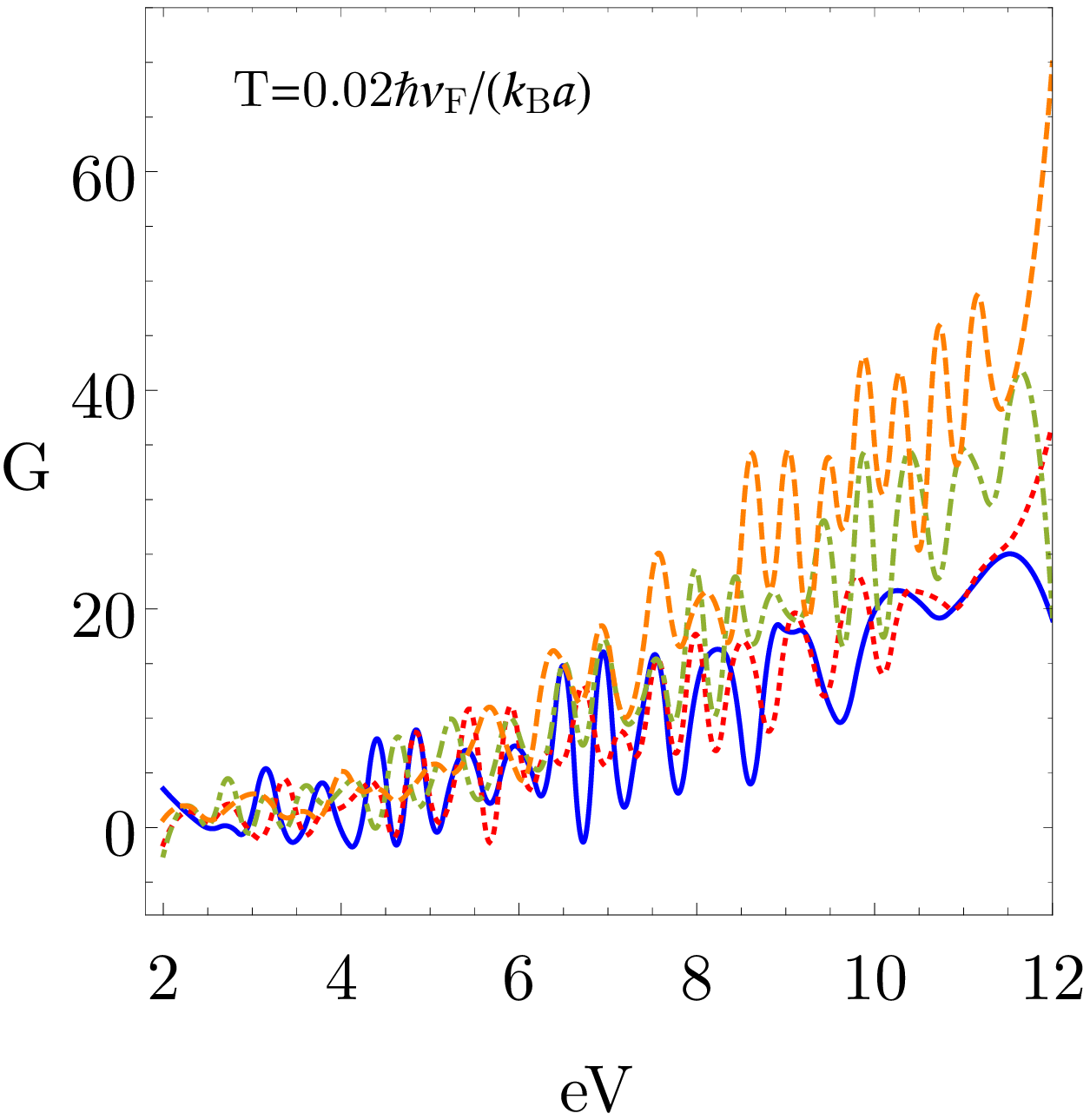}\label{fig6a}}
\subfigure[]{\includegraphics[width=0.3277\textwidth]{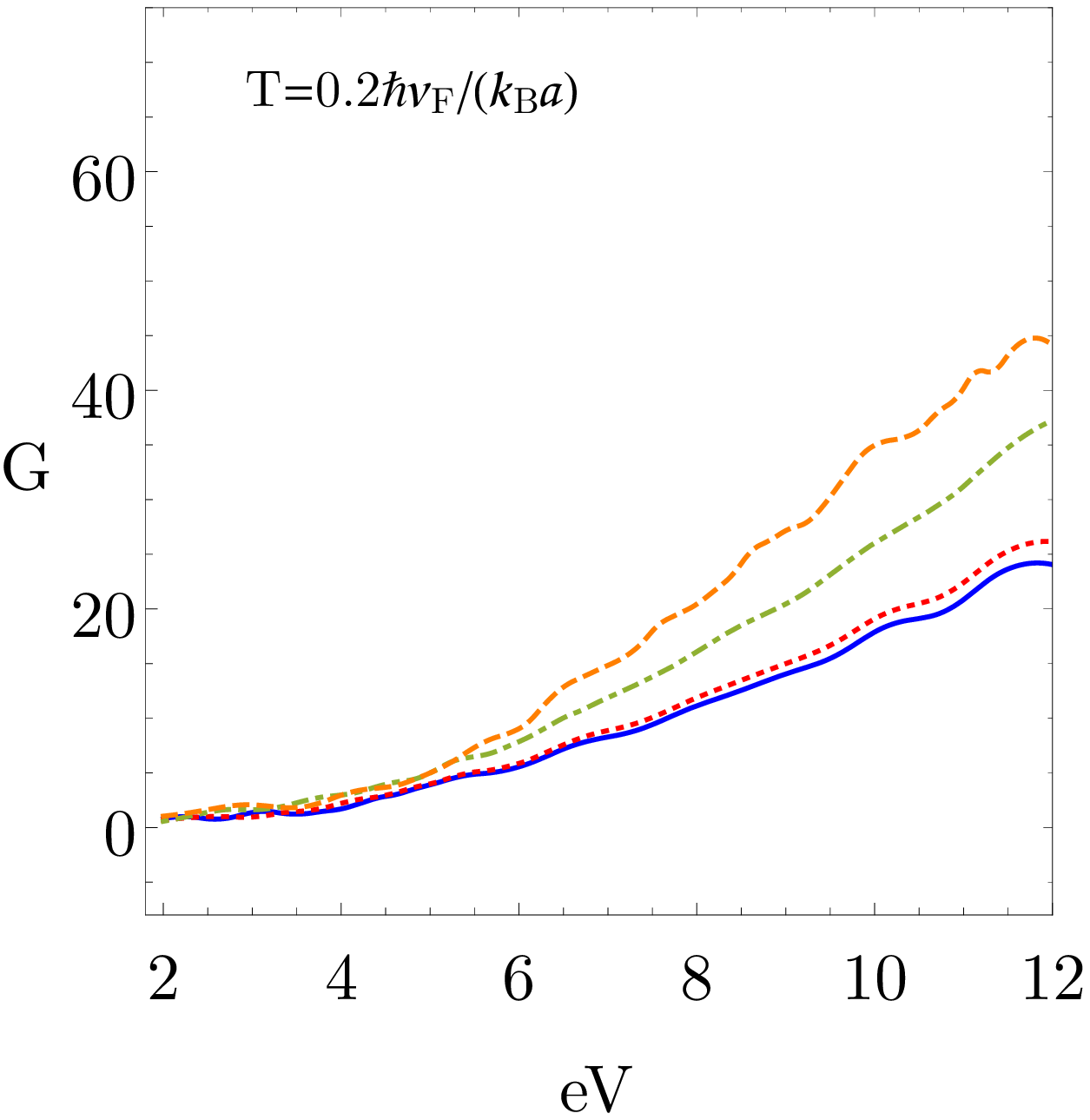}\label{fig6b}}
\caption{(Color online) Conductance (in units of $e^2/\hbar$) as a function of bias $e V$ (in units of $\hbar v_F/a$), calculated as the voltage-derivative of the analytical Eq.(\ref{eq_current}), for fixed $B_0a^2=2.4\tilde{\phi}_0$ and different values of $B_S$.
The solid (blue) line corresponds to $B_Sa^2=0$, the dotted (red) line corresponds to $B_Sa^2=0.5\tilde{\phi}_0$, the dotdashed (green) line corresponds to $B_Sa^2=1.1\tilde{\phi}_0$ and the dashed (orange) line corresponds to $B_Sa^2=1.7\tilde{\phi}_0$, with $\tilde{\phi}_{0}\equiv (v_F/c)\hbar /e$. The subfigures (a) and (b) correspond to the different values of the temperature $T = 0.02 \,\hbar v_F/ (k_B a)$ and $T = 0.2 \,\hbar v_F/ (k_B a)$ respectively.}
\label{fig6}
\end{figure}

\begin{figure}[hbt]
\centering
\includegraphics[width=0.3277\textwidth]{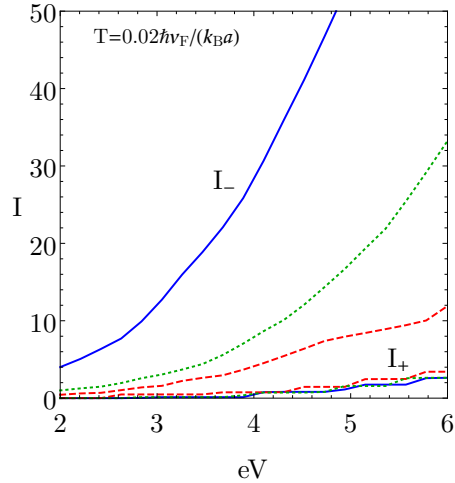}
\caption{(Color online) The valley-polarized components of the current (in units of $e v_F/a$), calculated from the analytical Eq.(\ref{eq_current}), are displayed as a function
of the applied bias $e V$ (in units of $\hbar v_F/a$), at finite temperature $T=0.02\,\hbar v_F/ (k_B a)$, fixed $B_0a^2=2.1\tilde{\phi}_0$ and different  values of $B_S$.
The solid (blue) lines correspond to $B_Sa^2=2\tilde{\phi}_0$, the dotted (green) lines correspond to $B_Sa^2=1.7\tilde{\phi}_0$ and the dashed (red) lines correspond to $B_Sa^2=1.1\tilde{\phi}_0$, with $\tilde{\phi}_{0}\equiv \hbar v_F/e$. The total current is $I = I_{+} + I_{-}$. A clear filtering effect in favour of the $I_{-}$ component versus the $I_{+}$ component is observed.}
\label{fig7}
\end{figure}

\begin{figure}[hbt]
\centering
\includegraphics[width=0.3277\textwidth]{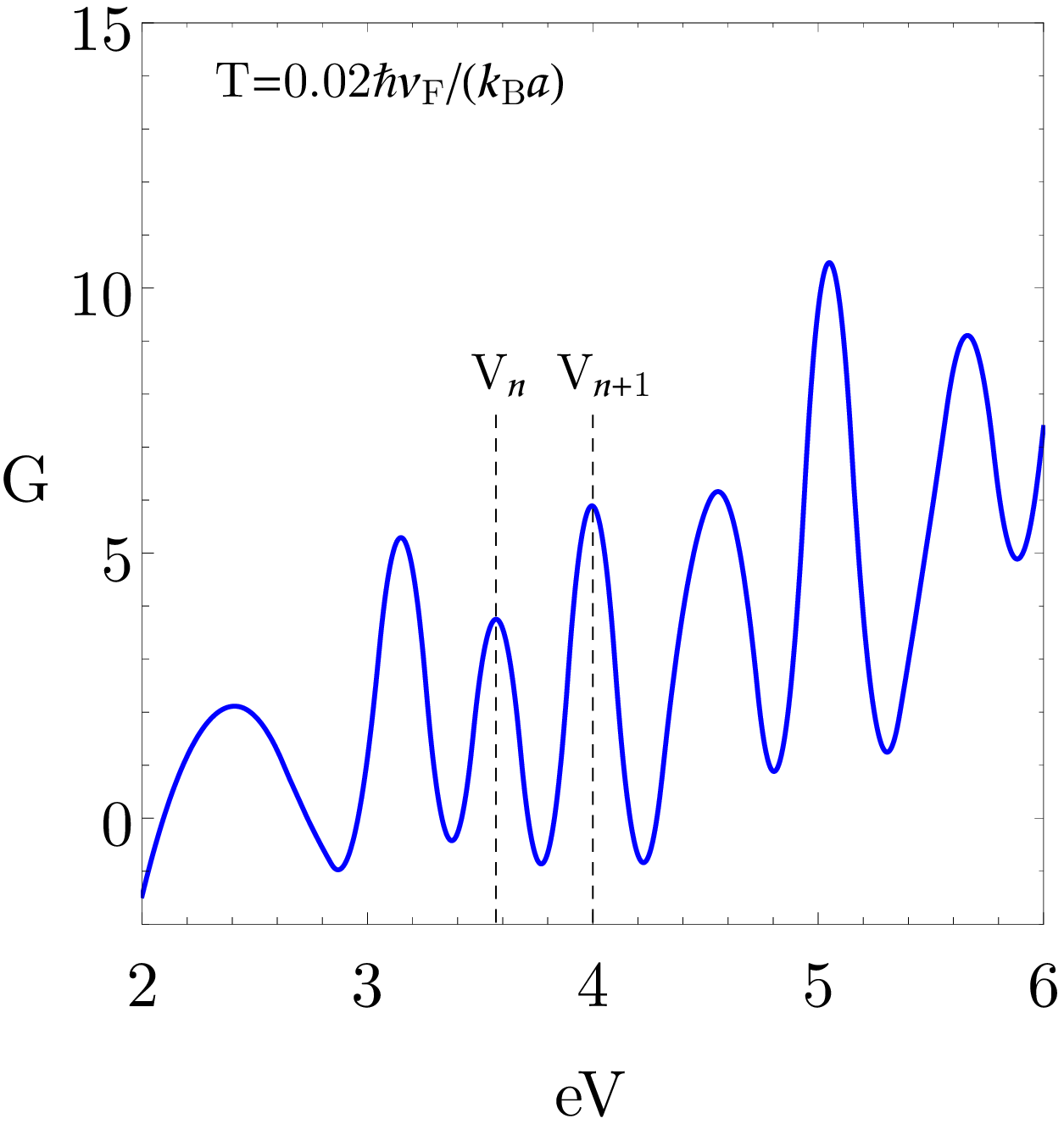}
\caption{(Color online) Conductance (in units of $e^2/\hbar$) as a function of bias $e V$ (in units of $\hbar v_F/a$), calculated as the voltage-derivative of the analytical Eq.(\ref{eq_current}), for fixed $B_0a^2=2.1\tilde{\phi}_0$ and $B_Sa^2=0.5\tilde{\phi}_0$ at a temperature $T = 0.02 \,\hbar v_F/ (k_B a)$}
\label{fig6_2}
\end{figure}

In this section, we represent graphically the total current $I = I_{+} + I_{-}$ (in units of $ev_F/a$) calculated from the analytical formula Eq.~(\ref{eq_current}) for the valley-polarized
components $I_{\xi}$, both at zero and
at finite temperatures (Figs. \ref{fig2} - \ref{fig4}), for the particular choice of contact chemical potentials $\mu_L = eV$ and $\mu_R = 0$. 
In Fig.\ref{fig2}(a) we represent the total current for $T = 0$, 
as a function of the applied bias voltage $V$, for a fixed value of the external magnetic field $B_0a^2 =1.8\tilde{\phi}_{0}$, with $\tilde{\phi}_{0}\equiv (v_F/c) \frac{\hbar}{e} $. The different curves display the dependence of the current on the magnitude of the
strain-induced pseudo-magnetic field $B_S$. Interestingly, there is a strong dependence of the current-voltage characteristics on the applied strain over the scattering region. One can also appreciate the staired shape of the curve, a feature
that is more manifest at lower values of strain. This effect follows directly from the condition of elastic
scattering (see Eq.\eqref{eq_transm}), since in order for the incident particle to be transmitted across the scattering region, its
incident energy must be resonant to one of the eigenstates in the circular region submitted to the fields. 
The quasi-continuum distribution of energy values in the reservoirs allows for this condition be always
fulfilled, for an interval within the window imposed by the external bias voltage. Moreover, let us notice that for contact chemical potentials $\mu_L = \alpha e V$ and $\mu_{R} = -\beta e V$, with $\alpha + \beta = 1$ such that
the net potential difference between the contacts is $V$, in the $T\rightarrow 0$ limit the difference between the two Fermi distributions in Eq.(\ref{eq_current}) becomes $f_L(E_{n}^{\xi}) - f_R(E_{n}^{\xi}) \rightarrow \Theta(\alpha eV - E_{n}^{\xi}) - \Theta(-\beta e V-E_{n}^{\xi})$. This difference vanishes outside the interval $E_{n}^{\xi} \in [-\beta eV,\alpha eV]$ that defines the energy window for allowed transmission, which clearly grows linearly with the bias voltage $V$, and hence more pseudo-Landau levels in the disk are resonant
for electronic transport as $V$ increases. Since these Landau levels are discrete, the transmission and correspondingly
the current increases by discrete steps, i.e. $E_n^{\xi} \sim \sqrt{|B_{\xi}|n}$. The distance between those steps decreases as the effective magnetic field diminishes, as actually occurs for the contribution
arising from the $\mathbf{K}_{-}$ cone, where $B_{-} = B_{0} - B_S$. As more and more discrete
Landau levels are included within the energy window imposed by the bias voltage, the current
increases accordingly. Hence, as clearly seen in Figs. \ref{fig2} - \ref{fig4}, the slope of the current-voltage characteristics, and hence the effective conductance across the region, increases
steadily with the magnitude of strain, for a fixed value of the external magnetic field $B_0$. It is worthwhile to analyze the effect of the voltage splitting parameters $\alpha$ and $\beta$. Let us first notice
that the distance between consecutive Landau levels is $E_{n+1}^{(\xi)} - E_{n}^{(\xi)} \sim \sqrt{2|B_{\xi}|}\left(\sqrt{n+1} - \sqrt{n} \right)\sim \frac{1}{\sqrt{n}}$, and hence the spectrum becomes denser as $n$ increases. Therefore,
a shift in the voltage window by choosing $\alpha < 1$ ($\alpha + \beta = 1$) will involve transmission of states from a less dense region of the spectrum (lower maximum value of $n$), thus decreasing the number
of channels (with respect to the case $\alpha=1$ and $\beta=0$) and hence the overall total current for the same net bias voltage $V$. The staircase pattern of the current-voltage characteristics clearly persists at finite temperature, but the steps are smeared, since the difference between the Fermi functions in Eq.(\ref{eq_current}) is smeared at finite temperatures and is no longer defined by the difference between two Heaviside functions, as mentioned before. This can be seen in the sub-figures(b)-(c) in Figs \ref{fig2} - \ref{fig4}, where the plateaus are smeared and even tend to disappear at high enough temperatures.

In Fig.\ref{fig3} at $B_{0} = 2.1\tilde{\phi}_0/a^2$ and Fig.\ref{fig4} at $B_{0} = 2.4\tilde{\phi}_0/a^2$, respectively, we compare the effect of the external magnetic field $B_0$ on the current-voltage characteristics. Both at $T=0$ and at $T>0$, it is seen that for the same values of strain, i.e. $B_S = (0.5 \tilde{\phi}_0/a^2, 1.1\tilde{\phi}_0/a^2, 1.7\tilde{\phi}_0/a^2)$, the total current decreases as the external magnetic field is increased from $B_0 = 1.8 \tilde{\phi}_0/a^2$ (in Fig.\ref{fig2}) towards $B_{0} = 2.4 \tilde{\phi}_0/a^2$ (in Fig.\ref{fig4}). This effect can be understood by the same argument presented before, since the density of the pseudo-Landau level spectrum increases
in the $\mathbf{K}_{-}$-valley as the magnitude of the effective pseudomagnetic field $|B_{-}| = |B_{S} - B_0|$ decreases. Thus, for a fixed strain field $B_S$, an increment in the external magnetic field $B_0$ leads to a reduction in $B_{-}$, with a subsequent increment of the spectral density that leads to an enhancement of the 
transmission and the corresponding $I_{-}$ component of the current.

The differential conductance $G(V,T) = dI/dV$ (in units of $e^2/\hbar$) at 
finite temperature $T = 0.02 \,\hbar v_F/ (k_B a)$ and $T = 0.2 \,\hbar v_F/ (k_B a)$ are displayed in Figs.~\ref{fig5} - \ref{fig6} for different values of $B_S$ and $B_0$. A characteristic trend of oscillations is observed, which are consistent with the staircase behavior of the current observed Figs.~\ref{fig2}--\ref{fig4}. Remarkably, this trend has also been measured experimentally in Ref.\cite{Yan_012}, where the linear slope of the valleys was attributed
 to a background voltage. However, our model explains the slope as a consequence of the finite temperature transmission mechanism since no external background is involved. Notice that the oscillations are strongly attenuated as the temperature increases, as shown in Fig.~\ref{fig5b} for  $T = 0.2 \,\hbar v_F/ (k_B a)$, due to the
 smearing effect of temperature over the Fermi-Dirac distribution in Eq.(\ref{eq_current}). It is also important to remark that, for a fixed value of the external magnetic field $B_0$, the differential conductance
 increases as increasing the strain field $B_S$, as clearly observed in the different curves represented in Fig.~\ref{fig5} and Fig.~\ref{fig6}. This effect can again be understood by
 noticing that the effective pseudomagnetic field $|B_{-}| = |B_{S} - B_0|$ decreases as $B_S$ increases, thus leading to a higher spectral density associated to the $\mathbf{K}_{-}$-valley and
 a corresponding enhancement of the transmission and conductance.

The relative enhancement of the valley-polarized contribution arising from the $\mathbf{K}_{-}$ valley is
clearly observed in Fig.\ref{fig7}, where the two valley components $I_{+}$ and $I_{-}$ of the total
current are represented at finite temperature. We notice that this effect is stronger when $B_S$ is closer to $B_0$, which is a consequence of the combination
of strain and a physical magnetic field, remains robust even at finite temperatures and hence may be used in practice to construct a valley-sensitive
filter.

On the other hand, the sensitivity of the current-voltage characteristics on the magnitude of strain, could be used in the construction of a nanoscale piezoelectric sensor based on graphene. The metrology principle
of the sensor can be based on determining experimentally the differential conductance $G = dI/dV$ for a fixed and controlled value of the
external magnetic field $B_0$, as displayed in Figs.~\ref{fig5a} and \ref{fig5b}, where the clearly distinguishable sharp peaks arise
as a consequence of each of the plateaus observed in the current-voltage characteristics. As discussed previously, the plateaus in the current, and hence the peaks in the conductance, arise from the bias voltage window imposed by two consecutive Landau levels (mainly from the $\mathbf{K}_{-}$-valley), i.e. $e\Delta V = eV_{n+1} - eV_{n} \sim E_{n+1}^{(-)} - E_{n}^{(-)} = \frac{\hbar v_F}{a}\sqrt{2|B_{-}|a^2/\tilde{\phi}_0}\left(\sqrt{n+1} - \sqrt{n}\right)$, where
in the equation we measure energies in units of $\hbar v_F/a$ and magnetic field in units of $\tilde{\phi}_0/a^2$. Therefore, by reading the
locus of two consecutive peaks $V_{n+1} > V_{n}$ in the conductance curve (see Fig.~\ref{fig6_2} for an example), it is possible to extract the value of the corresponding integer $n$ from the ratio:
\begin{eqnarray}
\frac{V_{n+1}}{V_n} \sim \frac{E_{n+1}^{(-)}}{E_{n}^{(-)}} = \sqrt{1 + \frac{1}{n}}\Longrightarrow n = \left\lfloor\frac{1}{\left(\frac{V_{n+1}}{V_n}\right)^{2}-1}\right\rceil.
\label{eq_voltage_peaks}
\end{eqnarray}
where the symbol $\lfloor x\rceil$ represents the nearest integer to $x$.
With the value of $n$, one can solve for the effective pseudo-magnetic field:
\begin{eqnarray}
|B_{-}| = \frac{\left(E_{n}^{(-)}/(\hbar v_F/a)\right)^2}{2 n}\frac{\tilde{\phi}_0}{a^2} \sim  \frac{\left(eV_n/(\hbar v_F/a)\right)^2}{2 n}\frac{\tilde{\phi}_0}{a^2} = \left|B_0 - B_S\right|.
\label{eq_B_extracted}
\end{eqnarray}
As a concrete example, let us take the values in Fig.~\ref{fig6_2}. We have that $n\approx\left\lfloor\frac{1}{\left(4.0/3.58\right)^{2}-1}\right\rceil=\lfloor4.02\rceil=4$ and from Eq. \eqref{eq_B_extracted} we have that $B_0 - B_S=3.58^2/(2\cdot 4) \tilde{\phi}_0/a^2=1.6\tilde{\phi}_0/a^2$. This gives exactly $B_{S}a^2=0.5\tilde{\phi}_0$, that was the value used to generate the conductance curve in  Fig.~\ref{fig6_2} in the first place. We can clearly see that this procedure can be applied in general and used to read off the effective strain magnetic field from the conductance curve. 
Since the conductance experiment is performed at a fixed and controlled value of the external field $B_0$, then the strain field $B_S$ is simply calculated
from Eq.(\ref{eq_B_extracted}). Under typical experimental conditions, the strain fields associated to graphene nanobubbles have been estimated on the order of $B_S \sim 100$ Tesla \cite{Levy_10,Peeters2013}. Therefore,
the externally imposed magnetic field satisfies $B_0 \ll B_S$, and hence no ambiguity in the sign of $B_\xi$ should arise in real experiments.

It is important to remark that typical experimental values for the characteristic diameter of graphene bubbles are $a \sim 15 - 25 nm$ \cite{Levy_10,Klimov_012, Yan_012,Zhu_015}, while
a graphene ribbon will have typical widths $W \sim 10 \mu m$ . Therefore, under realistic experimental conditions $a/W \ll 1$, and hence any influence of the edges of the ribbon over the carrier dynamics at the nanobubble becomes negligible. A possible exception is when the ribbon edges are saturated with $O$. This will lead to local magnetic moments that, under an externally imposed magnetic field $B_0$ as
described in our model, will tend to align parallel to the field, thus providing a small constant background field $B_{edge}$ that slightly modifies the one imposed externally, i.e. $B_{0}\rightarrow B_0 + B_{edge}$. If one
would like to take this correction into account in the estimation of $B_{S}$ from Eq.(\ref{eq_B_extracted}), the contribution of the magnetic moment at the edges can be calculated from simple stoichiometry by attributing a Bohr magneton $\mu_B$
unit to each magnetic moment at every Oxygen atom, or alternatively it can be obtained from an ab-initio calculation. In either case, the contribution will be very small compared to the magnetic field imposed for strain sensing purposes $B_{0}$, which should be
on the order of several Tesla.

\section{Conclusions and Summary}

In summary, we have provided a fully analytical treatment of a model for electronic transport through a graphene nanobubble, that combines the effects of
mechanical strain and an external magnetic field. Based on the partial wave analysis within scattering theory, we obtained analytical expressions for the transmission and valley-polarized
current components through the nanobubble, assuming that it is inmersed in a bulk graphene region connected to semi-infinite graphene contacts submitted
to different chemical potentials. Our analytical results predict a neat valley-polarization effect on the current, due to the combined effect of the local strain field and the externally imposed magnetic field, that determine the single-particle spectrum composed of pseudo-Landau levels. Moreover, we showed that the polarization effect is due to a valley-dependent enhancement of the spectral density through the pseudomagnetic fields $B_{\xi} = B_0 +\xi B_S$ at each valley $\mathbf{K}_{\xi}$. 

The predictions of this theoretical model, and particularly the sensitivity of the current-voltage characteristics on the magnitude of strain, could be used in the construction of a nanoscale piezoelectric sensor based on graphene. To measure strain patterns a the nanometer scale is experimentally difficult. 
However, as it was explained in detail in the previous section, our theoretical results suggest that by performing electronic conductance measurements the magnitude of such strain could in principle be inferred, giving a recipe for a strain-meter. The possibility of valley filtering, as it was previously suggested by numerical studies \cite{Settnes_016, Low_010}, it is here explicitly demonstrated with our completely analytical solution of the model.

\section*{Acknowledgements}

This work was supported by Fondecyt (Chile) Grant Nos. 1141146 (E. M.) and 11160542 (R. S.-G.)
\appendix

\section{Analytical solution of the 2D Dirac Hamiltonian under a constant magnetic field} 
\label{appendixA}

The general formulation of the eigenvalue problem for the Dirac equation in two dimensions, under a cylindrically symmetric magnetic field is presented. Moreover, the explicit
solution for the particular case of the uniform effective field arising from the combination of a magnetic field $\bm{B} = \hat{\bm e}_{3} B$, and a strain-induced pseudo-magnetic field at each Dirac cone $\mathbf{K}_{\pm}$, $\bm{B}_{S}^{\xi} = \hat{\bm e}_{3}\xi B_S$, such that $B_{\xi} = B + \xi B_{S}$. The notation and methods
are based on Ref.[\onlinecite{Thaller}]

Let us consider the case of a magnetic field normal to the plane of the sample, with cylindrical symmetry,
$\bm{B}_{\xi}(\bm{x}) = \hat{\bm e}_{3} B_{\xi}(r)$. The magnetic vector potential can be chosen in the gauge
$\bm{A} = \hat{\bm e}_{\phi} A_{\phi}(r)$, with
\begin{eqnarray}
A_{\phi}(r) = \frac{1}{r}\int_{0}^{r} B_{\xi}(\rho) \rho d\rho.
\label{eqA1}
\end{eqnarray}
Clearly, from Eq.(\ref{eqA1}) we have 
\begin{eqnarray}
B_{\xi}(r) = \frac{1}{r}\frac{d}{dr}(r A_{\phi}) = \frac{d}{dr}A_{\phi}(r) + \frac{1}{r}A_{\phi}(r).
\label{eqA2}
\end{eqnarray}

\subsection{The 2D Dirac Hamiltonian in cylindrical coordinates}

The Dirac Hamiltonian in 2D, in the presence of an external magnetic field, can be expressed by (in natural units $\hbar = 1$, $c=1$, $v_{F}\sim 1/300$)
\begin{eqnarray}
\hat{H}^{\xi}_S = \xi v_F \left(\vec{\sigma}\cdot\left(\bm{p} - \bm{A}\right) + M\hat{\sigma}_{3}\right),
\label{eqA3} 
\end{eqnarray}
where for the sake of completeness of the mathematical analysis, we have included the possibility of a finite mass $M$.
Let us define the unit vectors in cylindrical coordinates, $\hat{\bm r} = (\cos\phi,\sin\phi)$, $\hat{\bm \phi} = (-\sin\phi,\cos\phi)$. Then, the kinetic part of the Hamiltonian Eq.(\ref{eqA3}) can be expressed in cylindrical
coordinates as
\begin{equation}
\vec{\sigma}\cdot\left(\bm{p} - \bm{A}\right) = (\hat{\bm r}\cdot\vec{\sigma})(p_{r}-A_{r})
+ (\hat{\bm \phi}\cdot\vec{\sigma})(p_{\phi}-A_{\phi}).
\label{eqA4}
\end{equation}
Here, we have
\begin{equation}
\hat{\bm r}\cdot\vec{\sigma} = \cos\phi\,\hat{\sigma}_{1} + \sin\phi\,\hat{\sigma}_{2} = \left(
\begin{array}{cc}
0 & e^{-i\phi}\\e^{i\phi} & 0
\end{array}
\right),
\label{eqA5}
\end{equation}
and 
\begin{equation}
\hat{\bm \phi}\cdot\vec{\sigma} =  i(\hat{\bm r}\cdot\vec{\sigma})\hat{\sigma}_{3}.
\label{eqA6}
\end{equation}

Substituting into Eq.(\ref{eqA3}), and considering that $A_{r}=0$, we have
\begin{eqnarray}
\hat{H}_{S}^{\xi} = \xi v_F\left\{(\hat{\bm r}\cdot\vec{\sigma})\left[
-i\partial_{r} + i\hat{\sigma}_{3} \left(\frac{L_{3}}{r} - A_{\phi}\right)\right] + M\hat{\sigma}_{3}\right\}.
\nonumber\\
\label{eqA7}
\end{eqnarray}

Let us define the total angular momentum $\hat{J}_{3} = \hat{L}_{3} + \hat{\sigma}_{3}/2$. With this definition, the Hamiltonian in Eq.(\ref{eqA7}) becomes
\begin{eqnarray}
\hat{H}_S^{\xi} &=& \xi v_F\left\{(\hat{\bm r}\cdot\vec{\sigma})\left[
-i\left(\partial_{r} + \frac{1}{2 r}\right) + i\hat{\sigma}_{3} \left(\frac{\hat{J}_{3}}{r} - A_{\phi}\right)\right]+ M\hat{\sigma}_{3}\right\}.
\label{eqA8}
\end{eqnarray}

\subsection{Spinor eigenstates of $\hat{J}_{3}$}
It is straightforward to check that the two-component spinors
\begin{align}
\chi_{m_{j}} = \left(
\begin{array}{c} \chi_{1} e^{i(m_{j}-1/2)\phi}\\ \chi_{2} e^{i(m_{j} + 1/2)\phi}
\end{array}
\right),\quad & m_{j} = \pm \frac{1}{2},\pm \frac{3}{2},\ldots
\label{eqA9}
\end{align}
are eigenstates of $\hat{J}_{3}$ with eigenvalue $m_{j}$, i.e. $\hat{J}_{3}\chi_{m_{j}} = m_{j}\chi_{m_{j}}$. They also satisfy the property
\begin{align}
(\hat{\bm r}\cdot\vec{\sigma}) \chi_{m_{j}} =& \left(
\begin{array}{cc}
0 & e^{-i\phi}\\e^{i\phi} & 0\end{array}
\right) \left(
\begin{array}{c} \chi_{1} e^{i(m_{j}-1/2)\phi}\\ \chi_{2} e^{i(m_{j} + 1/2)\phi}
\end{array}
\right)=\left(
\begin{array}{c} \chi_{2} e^{i(m_{j}-1/2)\phi}\\ \chi_{1} e^{i(m_{j} + 1/2)\phi}
\end{array}
\right).
\label{eqA10}
\end{align}
Therefore, the eigenfunctions of the Hamiltonian Eq.(\ref{eqA8}) can be expressed, for given $m_{j}$, by the general form
\begin{equation}
\tilde{\Psi}_{m_{j}}^{\xi}(r,\phi) = \left(
\begin{array}{c}
\frac{1}{\sqrt{r}} f_{m_{j}}(r) e^{i(m_{j}-1/2)\phi}\\ -\frac{i}{\sqrt{r}}g_{m_{j}}(r)
e^{i(m_{j}+1/2)\phi}
\end{array}
\right)
\label{eqA11}
\end{equation}

\subsection{The eigenvalue equation}
The eigenvalue equation
\begin{equation}
\hat{H}_{S}^{\xi} \tilde{\Psi}_{m_{j}}^{\xi}(r,\phi) = E^{\xi} \tilde{\Psi}_{m_{j}}^{\xi}(r,\phi)
\label{eqA12}
\end{equation}
can be cast into the matrix operator form
\begin{equation}
\left[
\begin{array}{cc}
M & \hat{D}^{\dagger}\\ \hat{D} & -M
\end{array}
\right]\left(\begin{array}{c}
f_{m_{j}}\\g_{m_{j}}
\end{array}
 \right) \equiv \hat{h}_{m_{j}} \left(\begin{array}{c}
f_{m_{j}}\\g_{m_{j}}
\end{array}
 \right) = \xi v_F^{-1} E^{\xi} \left(\begin{array}{c}
f_{m_{j}}\\g_{m_{j}}
\end{array}
 \right).
 \label{eqA13}
\end{equation}
Here, we have defined the differential operator $\hat{D} = \frac{d}{dr} - \left(\frac{m_{j}}{r} - A_{\phi}  \right)$. Consequently, $\hat{D}^{\dagger} = -\frac{d}{dr} - \left(\frac{m_{j}}{r} - A_{\phi}  \right)$. From the algebraic
point of view, we have performed the subspace decomposition of the Hamiltonian operator:
$\hat{H}_S^{\xi} = \oplus_{m_{j}} \hat{h}_{m_{j}}$.
By direct calculation, one can show that the operators $\hat{D}$ and $\hat{D}^{\dagger}$ satisfy the relations
\begin{widetext}
\begin{eqnarray}
\hat{D}^{\dagger}\hat{D} &=& -\frac{d^{2}}{dr^{2}} + \frac{(m_{j}-1/2)^{2}-1/4}{r^{2}} -2\frac{(m_{j}-1/2)}{r} A_{\phi}
+ A_{\phi}^{2} - B_{\xi}(r),\nonumber\\
\hat{D}\hat{D}^{\dagger} &=& -\frac{d^{2}}{dr^{2}} + \frac{(m_{j}+1/2)^{2}-1/4}{r^{2}} -2\frac{(m_{j}+1/2)}{r} A_{\phi}
+ A_{\phi}^{2} + B_{\xi}(r).
\label{eqA14}
\end{eqnarray}
\end{widetext}
Here, we have used Eq.(\ref{eqA2}) for $B_{\xi}(r)$. Notice that it is more convenient to
square the effective Hamiltonian in the eigenvalue problem, to obtain the diagonal system 
\begin{equation}
\left[
\begin{array}{cc}
\hat{D}^{\dagger}\hat{D} + M^{2} - (\epsilon^{\xi})^{2} & 0\\ 0 & \hat{D} \hat{D}^{\dagger} + M^{2} - (\epsilon^{\xi})^{2}
\end{array}
\right]\left(
\begin{array}{c}
f\\g
\end{array}
\right) = 0,
\label{eqA15}
\end{equation}
where we defined $\epsilon^{\xi}=E^{\xi}/v_F$.

\subsection{Solution of the eigenvalue problem for uniform magnetic field}
Let us consider the particular case of a uniform magnetic field $B_{\xi}(r) = B_{\xi}$. Then, the magnetic vector potential Eq.(\ref{eqA1}) becomes $A_{\phi}(r) = B_{\xi} r /2$. Let us focus on the equation for $f(r)$, the upper component of the
spinor,
\begin{widetext}
\begin{eqnarray}
-\frac{d^{2}}{dr^{2}}f + \left[  \frac{(m_{j}-1/2)^{2}-1/4}{r^{2}} + \frac{B_{\xi}^{2}}{4}r^{2} \right] f
- \left[ \left(\frac{E^{\xi}}{v_F}\right)^{2} - M^{2} + B_{\xi} + (m_{j}-1/2)B_{\xi} \right] f = 0.
\label{eqA16}
\end{eqnarray}
\end{widetext}
We define the dimensionless variable $z = \frac{|B_{\xi}|}{2}r^{2} = \alpha r^{2}$, with $\alpha = |B_{\xi}|/2$. In terms of this new variable, Eq.(\ref{eqA16}) becomes
\begin{eqnarray}
z\frac{d^{2}f}{dz^{2}} + \frac{1}{2}\frac{df}{dz} - \left[ \frac{\kappa}{4 z} + \frac{z}{4} - \gamma \right]f = 0.
\label{eqA17}
\end{eqnarray}
Here, we have defined the parameters
\begin{eqnarray}
\gamma &=& \frac{\left(\frac{E^{\xi}}{v_F}\right)^{2} - M^{2} + (m_{j}-1/2)B_{\xi} + B_{\xi}}{4\alpha},\nonumber\\
\kappa &=& (m_{j}-1/2)^{2} - 1/4.
\label{eqA18}
\end{eqnarray}
Now, we analyze the asymptotic behaviour of $f(z)$ as $z\rightarrow 0$. If we write $f(z) \sim z^{q}$, then
substituting into Eq.(\ref{eqA17}) we see that it is satisfied for the leading terms as $z\rightarrow 0$ if
\begin{eqnarray}
q^{2} - \frac{q}{2} - \frac{\kappa}{4} = 0.
\label{eqA19}
\end{eqnarray}
The positive root of this equation is then $q = 1/4 + (1/2)\sqrt{1/4 + \kappa} = 1/4 + |m_{j}-1/2|/2$. 
On the other hand, as $z\rightarrow\infty$, the asymptotic form of Eq.(\ref{eqA17}) is
\begin{align}
z\frac{d^{2}f}{dz^{2}} - \frac{z}{4}f = 0,\,\,\, {\rm{as}}\,\,\,\, z\rightarrow\infty.
\label{eqA20}
\end{align}
This last expression possesses the asymptotic solution $f \sim e^{-z/2}$.
Considering the above expressions, we factor out the two asymptotic limits to write
\begin{align}
f(z) = z^{q} e^{-z/2} W(z).
\label{eqA21}
\end{align}
Inserting Eq.(\ref{eqA21}) into Eq.(\ref{eqA17}), and combining the definition of $\kappa$ in Eq.(\ref{eqA18}) with the value of $q$ obtained from Eq.(\ref{eqA19}), we have $q + 1/4 = 1/2 + |m_{j}-1/2|/2$, and $q^{2}-q/2-\kappa/4=0$. Hence, the differential Eq.(\ref{eqA17}) reduces to
\begin{widetext}
\begin{align}
z \frac{d^{2}W}{dz^{2}} + \left(1 + |m_{j}-1/2| - z \right) \frac{dW}{dz}
+ \left(
\gamma - \frac{1 + |m_{j}-1/2|}{2}
\right)W = 0,
\label{eqA22}
\end{align}
\end{widetext}
whose solutions are the Associated Laguerre polynomials \cite{Gradshteyn}
\begin{align}
W(z) = L_{n_{\rho}}^{|m_{j}-1/2|}(z), 
\label{eqA23}
\end{align}
provided the condition
\begin{align}
\gamma - \frac{1 + |m_{j}-1/2|}{2} = n_{\rho},\quad n_{\rho} = 0, 1, \ldots
\label{eqA24}
\end{align}
is satisfied. Combining Eq.(\ref{eqA18}) with the quantization condition Eq.(\ref{eqA24}), we solve
for the energy eigenvalues to be
\begin{align}
E_{\lambda}^{\xi}(n) = \lambda v_F \sqrt{2 n |B_{\xi}| + M^{2}},\,\,\,\,n = 0, 1, \ldots
\label{eqA25}
\end{align}
where $\lambda=\pm$ represents the ``band'' index, and we have defined 
\begin{align}
n = n_{\rho} + \frac{1}{2}\left(1  - {\rm{sgn}} B_{\xi} + |m_{j} - 1/2|- (m_{j} - 1/2){\rm{sgn}} B_{\xi}\right).
\label{eq_n}
\end{align}
We notice that the function $g_{m_j}(r)$ corresponding to the lower component of the spinor is not independent of the upper component
$f_{m_j}(r)$. Moreover, according to Eq.(\ref{eqA13}), $g_{m_j}(r)$ is given by the expression
\begin{eqnarray}
g_{m_j}(r) = \left(\xi v_F^{-1}E_{\lambda}^{\xi}(n)+ M \right)^{-1}\hat{D}f_{m_j}(r).
\label{eq_Ag1}
\end{eqnarray}
By considering separately the 4 different cases, i.e. $m_j-1/2 \ge \,\text{or}\,< 0$, and ${\rm{sgn}} B_{\xi} = \pm1$,
we obtain explicitly $g_{m_j}(r)$ and show that it corresponds to the same index $n$ defined in Eq.(\ref{eq_n}) and therefore corresponds to the same energy eigenvalue.
We shall use the following basic properties and recurrence relations for the Associated Laguerre polynomials \cite{Gradshteyn}
\begin{eqnarray}
\frac{d}{dz} L_{n}^{k}(z) &=& - L_{n-1}^{k+1}(z)= z^{-1}\left[ n L_n^{k}(z) - (n+k) L_{n-1}^k(z)  \right],
\label{eq_Lag1}
\end{eqnarray}
\begin{equation}
L_n^k(z) = L_n^{k+1}(z) - L_{n-1}^{k+1}(z),
\label{eq_Lag2}
\end{equation}
and combining Eq.(\ref{eq_Lag1}) and Eq.(\ref{eq_Lag2}), we obtain
\begin{eqnarray}
\frac{n}{z} L_{n}^{k-1}(z) = -L_{n-1}^{k+1}(z) + \frac{k}{z}L_{n-1}^{k}(z),
\label{eq_Lag3}
\end{eqnarray}
as well as
\begin{eqnarray}
\frac{n+1}{z}L_{n+1}^{k-1}(z) + L_n^k(z) = \frac{n + k}{z} L_n^{k+1}(z).
\label{eq_Lag4}
\end{eqnarray}

In terms of the dimensionless variable $z = |B_{\xi}|r^2/2$, we have $f_{m_j}(z) = z^{\frac{1}{4} + \frac{|m_j - 1/2|}{2}} e^{-z/2} L_{n_{\rho}}^{|m_j-1/2|}(z)$. In terms of this variable, the differential operator $\hat{D} = \sqrt{2|B_{\xi}|z} \left(\frac{d}{dz} -  \frac{m_j}{2 z} + \frac{{\rm{sgn}}B_{\xi}}{2}\right)$, and hence
Eq.(\ref{eq_Ag1}) becomes
\begin{eqnarray}
g_{m_j}(z) &=& \frac{\sqrt{2|B_{\xi}|}}{\xi v_F^{-1}E_{\lambda}^{\xi}(n)+ M }\, z^{\frac{3}{4} + \frac{|m_j - 1/2|}{2}} e^{-z/2}\nonumber\\
&&\times\left(\frac{d}{dz} L_{n_{\rho}}^{|m_j - 1/2|}(z)
+ \left[ \frac{|m_j - 1/2| - (m_j - 1/2)}{2 z}+ \frac{{\rm{sgn}}B_{\xi} - 1}{2} \right] L_{n_{\rho}}^{|m_j - 1/2|}(z)\right).
\label{eq_Ag2}
\end{eqnarray}
Let us now reduce Eq.(\ref{eq_Ag2}) to the minimal expression, by considering the 4 separate cases:
\subsubsection{Case 1: $m_j - 1/2 \ge 0$, and ${\rm{sgn}}B_{\xi} = +1$}
In this case, by using the first identity in Eq.(\ref{eq_Lag1}) we have $\frac{d}{dz} L_{n_{\rho}}^{m_j - 1/2}(z) = - L_{n_{\rho}-1}^{m_j + 1/2}(z)$, and hence Eq.(\ref{eq_Ag2}) reduces to
\begin{eqnarray}
g_{m_j}(z) = - \frac{\sqrt{2|B_{\xi}|}}{\xi v_F^{-1}E_{\lambda}^{\xi}(n)+ M }\, z^{\frac{3}{4} + \frac{m_j-1/2}{2}} e^{-z/2} L_{n_{\rho}-1}^{m_j + 1/2}(z).\nonumber\\
\label{eq_gc1}
\end{eqnarray}

\subsubsection{Case 2: $m_j - 1/2 \ge 0$ and ${\rm{sgn}}B_{\xi} = -1$}
In this case, we use identities Eq.(\ref{eq_Lag1}) and Eq.(\ref{eq_Lag2}) as follows
\begin{eqnarray}
\frac{d}{dz}L_{n_{\rho}}^{m_j - 1/2}(z) - L_{n_{\rho}}^{m_j - 1/2}(z) &=& - L_{n_{\rho}-1}^{m_j+1/2}(z) - L_{n_{\rho}}^{m_j - 1/2}(z)= - L_{n_{\rho}}^{m_j + 1/2}(z).
\end{eqnarray}
Therefore, for this case Eq.(\ref{eq_Ag2}) reduces to the expression
\begin{eqnarray}
g_{m_j}(z) = - \frac{\sqrt{2|B_{\xi}|}}{\xi v_F^{-1}E_{\lambda}^{\xi}(n)+ M }\, z^{\frac{3}{4} + \frac{m_j - 1/2}{2}} e^{-z/2} L_{n_{\rho}}^{m_j + 1/2}(z).\nonumber\\
\label{eq_gc2}
\end{eqnarray}

We further notice that, in terms of the index that defines the energy eigenvalue, we have for $m \equiv m_j - 1/2 \ge 0$ after Eq.(\ref{eq_n}) $n_{\rho} = n - (1+m)\frac{1-{\rm{sgn}}B_{\xi}}{2}$. Substituting into Eq.(\ref{eqA11}), we have that the full spinor
eigenfunction near the cone $K_{\xi}$, with energy eigenvalue $E_{\lambda}^{\xi}(n)$, for $m \equiv m_j - 1/2 \ge 0$ and $n > 0$ 
\begin{align}
&\tilde{\Psi}_{n,m}^{\xi,\lambda}(r,\phi) = C_{m,n}^{\xi,\lambda}\left(\begin{array}{cc}z^{\frac{m}{2}} e^{-z/2} L_{n - (1+m)\theta(-B_{\xi})}^{m}(z)e^{i m\phi}\\\frac{i\sqrt{2|B_{\xi}|}\,z^{\frac{m+1}{2}}e^{-z/2}}{\lambda\xi \sqrt{2 n |B_{\xi}| + M^2}+M} L_{n - 1-m\theta(-B_{\xi})}^{m+1}(z) e^{i (m+1)\phi} \end{array}\right).
\end{align}

Here, $\theta(x)$ is the Heaviside step function.

\subsubsection{Case 3: $m_j < 1/2$ and ${\rm{sgn}}B_{\xi} = +1$}
Here, it is convenient to write $m_j - 1/2 = -|m_j - 1/2|$, and hence using Eq.(\ref{eq_Lag1})
\begin{align}
&\frac{d}{dz} L_{n_{\rho}}^{|m_j - 1/2|}(z) + \frac{|m_j-1/2|}{z}L_{n_{\rho}}^{|m_j - 1/2|}(z)\nonumber\\ &= - L_{n_{\rho}-1}^{|m_j-1/2| + 1}(z) + \frac{|m_j-1/2|}{z}
L_{n_{\rho}}^{|m_j - 1/2|}(z)\nonumber\\
&= L_{n_{\rho}}^{|m_j-1/2|}(z) - L_{n_{\rho}}^{|m_j-1/2|+1}(z) + \frac{|m_j-1/2|}{z}L_{n_{\rho}}^{|m_j-1/2|}(z)\nonumber\\
&= L_{n_{\rho}}^{|m_j-1/2|}(z) + \frac{n_{\rho}+1}{z} L_{n_{\rho}+1}^{|m_j-1/2|-1}(z)\nonumber\\
&= \frac{n_{\rho}+ |m_j-1/2|}{z} L_{n_{\rho}}^{|m_j-1/2|+1}(z)\nonumber\\
&= \frac{n_{\rho}+ |m_j-1/2|}{z} L_{n_{\rho}}^{|m_j+1/2|}(z).
\end{align}
where in the second step we used Eq.(\ref{eq_Lag3}). Thus, for this case we finally obtain
\begin{align}
g_{m_j}(z) &= - \frac{\sqrt{2|B_{\xi}|}(n_{\rho}+|m_j-1/2|)}{\xi v_F^{-1}E_{\lambda}^{\xi}(n)+ M }\, z^{\frac{1}{4}- \frac{m_j + 1/2}{2}} e^{-z/2} L_{n_{\rho}}^{|m_j + 1/2|}(z).
\label{eq_gc3}
\end{align}

\subsubsection{Case 4: $m_j-1/2 < 0$ and ${\rm{sgn}}B_{\xi} = -1$}
For this case, we use again Eq.(\ref{eq_Lag1}) to reduce
\begin{align}
&\frac{d}{dz}L_{n_{\rho}}^{|m_j - 1/2|}(z) + \left(\frac{|m_j-1/2|}{2z} -1 \right)L_{n_{\rho}}^{|m_j - 1/2|}(z)\nonumber\\
&= - \left(L_{n_{\rho}-1}^{|m_j - 1/2|+1}(z) + L_{n_{\rho}}^{|m_j-1/2|}(z)\right)+ \frac{|m_j - 1/2|}{z} L_{n_{\rho}}^{|m_j-1/2|}(z)\nonumber\\
& = - L_{n_{\rho}}^{|m_j - 1/2|+1}(z) + \frac{|m_j - 1/2|}{z} L_{n_{\rho}}^{|m_j-1/2|}(z)\nonumber\\
& = \frac{n_{\rho}+1}{z} L_{n_{\rho}+1}^{|m_{j}-1/2|-1}(z)\nonumber\\
& = \frac{n_{\rho}+1}{z} L_{n_{\rho}+1}^{|m_{j}+1/2|}(z).
\end{align}
where in the second step we used the identity Eq.(\ref{eq_Lag2}), and in the last step we used Eq.(\ref{eq_Lag4}).

Therefore, for this case we obtain the final expression
\begin{eqnarray}
g_{m_j}(z) = - \frac{\sqrt{2|B_{\xi}|}(n_{\rho}+1)}{\xi v_F^{-1}E_{\lambda}^{\xi}(n)+ M }\, z^{\frac{1}{4} - \frac{m_j + 1/2}{2}} e^{-z/2} L_{n_{\rho}+1}^{|m_j + 1/2|}(z).\nonumber\\
\label{eq_gc4}
\end{eqnarray}

In conclusion,
for the case $m_j - 1/2 \equiv m < 0$, the spinor eigenvector near the cone $K_{\xi}$, with energy eigenvalue $E_{\lambda}^{\xi}(n)$, is given by
\begin{align}
&\tilde{\Psi}_{n,m}^{\xi,\lambda}(r,\phi) = C_{m,n}^{\xi,\lambda}\left(\begin{array}{cc}z^{\frac{|m|}{2}} e^{-z/2} L_{n-\theta(-B_{\xi}) - |m| \theta(B_{\xi})}^{|m|}(z)e^{i m\phi}\\\frac{i\sqrt{2|B_{\xi}|}n\,z^{\frac{|m+1|}{2}}e^{-z/2}}{\lambda\xi \sqrt{2 n |B_{\xi}| + M^2}+M} L_{n-|m|\theta(B_{\xi})}^{|m+1|}(z) e^{i (m+1)\phi} \end{array}\right).
\end{align}

To summarize the results, and in order to calculate the normalization coefficient, we have the general solution (for $n>0$)
\begin{align}
&\tilde{\Psi}_{n,m}^{\xi,\lambda}(r,\phi) = C_{m,n}^{\xi,\lambda}\left(\begin{array}{cc}z^{\frac{|m|}{2}} e^{-z/2} L_{n_{\rho}}^{|m|}(z)e^{i m\phi}\\i\,\alpha_{n}^{\xi}z^{\frac{|m+1|}{2}}e^{-z/2} L_{n'_{\rho}}^{|m+1|}(z) e^{i (m+1)\phi} \end{array}\right).
\end{align}
Here, the coefficients are defined by
\begin{eqnarray}
n_{\rho} &=& n - \theta(-B_{\xi}) - \frac{|m| - m\,{\rm{\sgn}}B_{\xi}}{2},\nonumber\\
n'_{\rho} &=&  n_{\rho} - \theta(B_{\xi}) + \theta(-m),\nonumber\\
\alpha_{n}^{\xi} &=& \frac{\sqrt{2 |B_{\xi}| }n^{\theta(-m)}}{\lambda\xi \sqrt{2 n |B_{\xi}| + M^2} + M}.
\end{eqnarray}
Here, $C_{m,n}^{\xi,\lambda}$ is a normalization constant, chosen such that each eigenvector has unit norm, as follows ($z = |B_{\xi}|r^2/2$)
\begin{align}
1 =& \frac{1}{|B_{\xi}|}\int_{0}^{2\pi} d\phi \int_{0}^{\infty} dz\,|\Psi_{n_{\rho},m_{j}}^{\lambda}(z,\phi)|^2\nonumber\\
=& \frac{|C_{m,n}^{\xi,\lambda}|^{2}}{|B_{\xi}|}\, 2\pi \left\{
\int_{0}^{\infty} dz z^{|m|} e^{-z} 
\left[ L_{n_{\rho}}^{|m|}(z) \right]^{2}+ \left(\alpha_n^{\xi}\right)^2 \int_{0}^{\infty} dz z^{|m+1|} e^{-z}\left[ L_{n'_{\rho}}^{|m+1|}(z) \right]^{2}\right\} .
\label{eqA31}
\end{align}
Using the identity (Gradshteyn, 7.414-3)
\begin{equation}
\int_{0}^{\infty} e^{-x} x^{\alpha} L_{n}^{\alpha}(x) L_{m}^{\alpha}(x) = \delta_{n,m} \frac{\Gamma(\alpha+n+1)}{n!},
\label{eqA36}
\end{equation}
and solving Eq.(\ref{eqA31}), we find the final expression (for $n>0$)
\begin{align}
C_{m,n}^{\xi,\lambda} &= \left( \frac{|B_{\xi}|}{2\pi} \right)^{\frac{1}{2}}
\left\{ 
\frac{\Gamma(|m| + n_{\rho} + 1)}{n_{\rho}!}+ \left(\alpha_n^{\xi} \right)^2\frac{\Gamma(|m+1| + n'_{\rho} + 1)}{n'_{\rho}!} \right\}^{-1/2}.
\label{eqA37}
\end{align}
The state with $n = 0$ must be analyzed separately. For $\sgn B_{\xi} = +1$, following the steps of case 1 above, that the state $n = 0$
is only compatible with $m \ge 0$, where $L_{0}^{m}(z) = 1$. Therefore, we have for $\sgn B_{\xi} = +1$
\begin{equation}
\tilde{\Psi}_{0,m\ge0}^{\xi,\lambda}(z,\phi) = C_{m\ge0,0}^{\xi,\lambda}\left(\begin{array}{c} z^{\frac{m}{2}} e^{-z/2} e^{i m \phi}\\ 0 \end{array} \right).
\end{equation}
On the other hand, for $\sgn B_{\xi} = -1$, following the steps of case 4 above, we find that the state $n=0$ is only compatible with $m < 0$. Therefore, we have for $\sgn B_{\xi} = -1$
\begin{equation}
\tilde{\Psi}_{0,m<0}^{\xi,\lambda}(z,\phi) = C_{m<0,0}^{\xi,\lambda}\left(\begin{array}{c}0\\ z^{\frac{|m+1|}{2}} e^{-z/2} e^{i (m+1) \phi} \end{array} \right).
\end{equation}
Here, the normalization coefficients are given by
\begin{align}
C_{m,0}^{\xi,\lambda} &=\left( \frac{|B_{\xi}|}{2\pi}\right)^{1/2}\left\{
\theta (B_{\xi}) \Gamma(|m| + 1)+ \theta(-B_{\xi})\Gamma(|m+1| + 1)
\right\}^{-1/2}.
\end{align}

%


\end{document}